\documentclass[final,5p,times,twocolumn,number]{elsarticle}

\usepackage{amssymb}
\usepackage{amsmath}
\usepackage{lipsum}
\usepackage{graphicx}
\usepackage{subcaption}
\usepackage{float}

\journal{Nuclear Physics B}
\makeatletter
\def\ps@pprintTitle{%
 \let\@oddhead\@empty
 \let\@evenhead\@empty
 \def\@oddfoot{\footnotesize\itshape{} \hfill\today}%
 \let\@evenfoot\@oddfoot}
\makeatother

\begin{document}

\begin{frontmatter}

\title{Restricted Phase Space Thermodynamics of 4D Dyonic AdS Black Holes: Insights from Kaniadakis Statistics and Emergence of Superfluid $\lambda$-Phase Transition}

\author[first]{Abhishek Baruah}
\author[second]{Prabwal Phukon$^{a,}$}

\address[first]{Department of Physics, Dibrugarh University, Dibrugarh, Assam 786004, India}
\address[second]{Theoretical Physics Division, Centre for Atmospheric Studies, Dibrugarh University, Dibrugarh, Assam 786004, India}

\begin{abstract}
We study the thermodynamics of $4D$ dyonic AdS black hole in the Kaniadakis statistics framework using the Restricted Phase Space (RPST) formalism. This framework provides a non-extensive extension of classical statistical mechanics, drawing inspiration from relativistic symmetries and presenting a fresh perspective on black hole thermodynamics. Our study analyzes how including Kaniadakis entropy modifies the phase transition of the dyonic black holes. We consider the central charge $C$ and its conjugate chemical potential $\mu$ as the thermodynamic variable along with others except the pressure and volume. Due to the addition of the magnetic charge $\tilde{Q}_m$, the study of the phase transition becomes much richer by obtaining a non-equilibrium phase transition from an unstable small black hole to a stable large black hole along with the Van der Waals phase transition in the $T-S$ processes. In the $F-T$ plot, we get an extra Hawking-Page phase transition. Including the deformation parameter $\kappa$ introduces an unstable (ultra-large BH) branch seen in almost all the plots. Turning off the magnetic charge flips the direction of the phase transition seen during its presence. We observe a novel phenomenon that is the superfluid $\lambda$ phase transition in the mixed $(\tilde{\Phi}_e,\tilde{Q}_m)$ which is due to the additional $\tilde{Q}_m$ inclusion. Also, in the plots varying $\kappa$ match with the plot varying $C$ which underlines some sort of correspondence in its meaning which is not possible to observe in Gibbs-Boltzmann statistics. As the entropy models change the homogeneity is not lost where mass is of the first order and the rest is zeroth order. Finally, the $\mu-C$ processes in quite similar across black hole systems and entropy formulation marking some kind of universality of this process.
\end{abstract}


\end{frontmatter}




\section{Introduction}
\label{sec:introduction}
The exploration and study of black hole thermodynamics has performed a pivotal role in bridging classical gravity, quantum mechanics, and statistical mechanics offering a profound framework for understanding the microscopic nature of spacetime. This study began with the introduction of Hawking's temperature $T$ and Bekenstein-Hawking entropy given as:
\begin{equation}
    T=\frac{\hbar \kappa}{2 \pi ck_B }, \quad S=\frac{k_B c^3 A}{4 \hbar G}
\end{equation}
where $\kappa$ is the surface gravity and $A$ is the horizon area \cite{a,ak1,b,bb1}. It has led to an active area of research \cite{0a,0b,0c,0d,0e}. AdS black holes hold a unique place in this discourse due to their thermodynamic property as seen by Hawking and Page as they observed a phase transition between a pure thermal AdS space and the Schwarzschild AdS black hole \cite{d}. There are even more ways to study the various phase transitions \cite{1a,1b,1c,1d,1e,1f} namely via the thermodynamic geometry \cite{2a,2b,2c,2d,2e,2f,2g,2h,2i,2j,2k} and topological studies \cite{3a,3b,3c,3d,3e,3f,3g,3h,3i,3j,3k,3l,3m,3n,3o,3p,3q,3r,3s,3t,3u,3v,3w,3x} .\\
A revolutionary leap made in the study of black hole thermodynamics in AdS spacetimes is the Extended Phase Space Thermodynamics (EPST) formalism by incorporating the $(P,V)$ pair as variables where $P=-\Lambda/8 \pi G$. This approach led to much research particularly focusing on $P-V$ criticality \cite{i,x,y,yy,yy1,yy2,yy3,yy4,yy5,yy6} also enabling it to be considered as heat engines \cite{b1,bc1} With this black hole was considered to have duality with other systems like QCD, CMP, and CFT \cite{f,g,h,j,k,l,m,n,o,o1,o2,j1}. Visser proposed further advancement in this framework \cite{u}, who was inspired by the AdS/CFT correspondence \cite{c} and introduced a new parameter the central charge $C$ and its conjugate chemical potential $\mu$ as variables replacing the $(P-V)$ with CFT-inspired $(\mathcal{P}-\mathcal{V})$ where $V \sim L^{d-2}$ where $L$ is the AdS radius and $\mathcal{P}$ determined by the EOS $E=(d-2)\mathcal{P}\mathcal{V}$ where $d$ is the bulk spacetime dimension \cite{p,v,vv}.  This inclusion restored the interpretation of total black hole mass as internal energy, aligning with the traditional thermodynamic principles. For a charged rotating AdS black hole, the first law is given as:-
\begin{equation}
    dE=TdS-\mathcal{P}d\mathcal{V}+\tilde{\Phi}d \tilde{Q}+\Omega dJ+\mu C
\end{equation}
and the Smarr relation is given as:-
\begin{equation}
    E=TS+\tilde{\Phi}\tilde{Q}+\Omega J+\mu C
\end{equation}
where $\tilde{\Phi}$ and $\tilde{Q}$ are rescaled potentials and charges.
Mostly, Visser's work unified the bulk and boundary thermodynamics, offering a base to describe the black hole and its holographic CFT dual. Despite the tremendous advancement, the EPS and Visser's formalism faced certain challenges which included issues like ensemble ambiguity due to cosmological variation linked to the central charge and homogeneity of the energy function \cite{z,a1,c1,d1,e1,f1,f11,f12,f13}. To overcome and address these concerns, the Restricted Phase Space (RPST) formalism was introduced \cite{rps} as a refined alternative by fixing the central charge $C$. Via this approach, the ambiguity is avoided and a consistent thermodynamic description is maintained allowing various studies to be taken into consideration. Many studies has been conducted using RPS formalism \cite{rps1,rps2,rps3,rps4,rps5,rps6,rps7,rps8}.\\
Dyonic AdS black holes are characterized by the possession of electric and magnetic charges simultaneously creating a compelling class of solutions in the four-dimensional spacetimes. Due to the electromagnetic duality, these black holes offer a unique platform for exploring thermodynamic properties. Considering dyonic black holes in asymptotically AdS spacetime, we can study the behavior of the dual field theories \cite{l2}, shedding light on condensed matter physics. The dual field theory on the boundary is deformed by the black hole's $U(1)$ gauge field supporting two distinct modes: a normalizable mode relating to the vacuum expectation value and a non-normalizable mode that relates to an external gauge field influencing the boundary. Due to this deformation, various condensed matter effects are understood such as superconductivity/superfluidity, the Hall effect, magnetohydrodynamics, and the Nernst effect \cite{aa,aa1,ab,ac,ac1,ac2,ac3,ac4,ac5}. \\
One of the most universal and fundamental concepts in physics has been entropy since the advent of classical thermodynamics. But non-extensive thermodynamics has revealed that entropy is not as fundamental or unique showing context-dependent which evolves across theories enabling re-evaluation of the principles underlying the definition and the role of entropy \cite{en1}. Bekenstein seminal work being associated with the entropy of black holes proposed that it is proportional to the horizon rather than the volume, a major deviation from the classical thermodynamics where entropy is proportional to the system's mass and volume and is extensive. The discovery of Hawking's radiation that black holes emit a black body spectrum with a temperature $T$, established the Bekenstein-Hawking entropy as a main block of black hole thermodynamics.\\ Despite its success the non-extensiveness of Bekenstein-Hawking entropy remains enigmatic because classical thermodynamic entropy is additive and extensive but black hole entropy defies this principle signaling interest in alternate formulations \cite{en01,en02,en03,en04,en05,en2,en3,en4,en5,en6,en7,en8,en9,en10,en11,en12,en13,en14,en15,en16,en17,en18,en19}. Non-additive entropy models such as R\'enyi  and Tsallis entropies \cite{en20,en21} provide new insights into systems characterized by non-extensive behaviors. Recent developments also include Barrow entropy \cite{en22}, Sharma Mittal \cite{en23} and Kaniadakis entropy proposals \cite{en24,en25} which aim to address the limitations of the classical formulations. The Kaniadakis entropy is given as:-
\begin{equation}
     S_K= \frac{1}{\kappa} \sinh{(\kappa S_{BH})}
\end{equation}
where $\kappa$ is the deformtion parameter and when $\kappa \rightarrow{0}$, it reproduces the Bekenstein-Hawking entropy. \\
Building on the foundation of restricted phase space thermodynamics (RPST) using Bekenstein-Hawking and Rényi entropies \cite{me}, this paper extends the study to the thermodynamics of dyonic AdS black holes by incorporating Kaniadakis entropy. The inclusion of magnetic charge introduces a new thermodynamic ensemble $(\tilde{\Phi}_e,\tilde{Q}_m)$, enriching the phase structure with intricate interactions. Kaniadakis entropy, known for its non-extensive statistical mechanics framework, introduces an additional unstable branch in the black hole phase structure, presenting novel and intriguing phenomena. This work aims to explore these effects comprehensively within the RPST framework, offering fresh insights into the interplay between non-extensive statistics and black hole thermodynamics, thereby advancing our understanding of the underlying physics.\\
This paper is organized as follows: In Section \ref{sec:introduction} we give a brief introduction about Restricted Phase Space (RPS) formalism, dyonic AdS black holes, and the idea of implementing Kaniadakis statistics and also the motivation of our work carried out. In Section \ref{sec:RPST}, we show the RPS thermodynamics of dyonic AdS black holes. In Section \ref{sec:EOS} we show the homogeneity using the Kaniadakis entropy and also study the various processes namely the $T-S$, $F-T$ $\tilde{\Phi}_e-\tilde{Q}_e$, $\tilde{\Phi}_m-\tilde{Q}_m$ and $\mu-C$ processes. In Section \ref{sec:Discussions} we discuss certain important points which we understand from the paper and what this study offered us.

\section{RPST thermodynamics of dual charged black holes}
\label{sec:RPST}
The Reissner-Nordstr$\ddot{o}$m action in 4-D is given as:-
\begin{equation}
    I=\frac{1}{16 \pi G_4}\int d^4x\sqrt{g}(-R+F^2-\frac{6}{b^2})
\end{equation}
We write the equation of motion as:-
\begin{equation}
    R_{\mu \nu}-\frac{1}{2}g_{\mu\nu}R-\frac{3}{b^2}g_{\mu\nu}=2(F_{\mu\lambda}F^\lambda_\nu-\frac{1}{4}g_{\mu\nu}F_{\alpha\beta}F^{\alpha\beta})
\end{equation}
\begin{equation}
    \Delta_\mu F^{\mu\nu}=0
\end{equation}
A static symmetric solution is:-
\begin{equation}
    A=\left(\frac{-Q_e}{r}+\frac{Q_e}{r_+}\right)dt+(Q_m \cos\theta)d\phi
\end{equation}
The metric is obtained as:-
\begin{equation}
    ds^2=-f(r)dt^2+\frac{1}{f(r)}dr^2+r^2d\theta^2+r^2\sin^2\theta^2d\phi^2
\end{equation}
where $f(r)=1+\frac{r^2}{l^2}-\frac{2GM}{r}+\frac{G(Q_e^2+Q_m^2)}{r^2}$, where $Q_e$ is the electric charge, $Q_m$ is the magnetic charge $M$ is the mass of the black hole and $r$ is the horizon radius. Setting $f(r_+)=0$, the mass is:-
\begin{equation}
    \label{eq:mass1}  M=\frac{Gl^2Q_e^2+Gl^2Q_m^2+l^2r^2+r^4}{2Gl^2r}
\end{equation}
We add a rescaled electric charge $\tilde{Q}_e=\frac{Q_e l}{\sqrt{G}}$, electric potential $\tilde{\Phi}_e=\frac{\Phi_e\sqrt{G}}{l}$, magnetic charge $\tilde{Q}_m=\frac{Q_ml}{\sqrt{G}}$ and magnetic potential $\tilde{\Phi}_m=\frac{\Phi\sqrt{G}}{l}$. Introducing the central charge $C$ and then the chemical potential $\mu$ is:-
\begin{equation}
\label{eq:centralcharge}
    C=\frac{l^2}{G}, \quad \mu=\frac{M-TS-\tilde{\Phi}_e\tilde{Q}_e-\tilde{\Phi}_m\tilde{Q}_m}{C}
\end{equation}
The central charge is defined as the effective number of the microscopic degrees of freedom $N_{bulk}$ and its conjugate $\mu$ as the chemical potential of the black hole in the bulk. To serve a connection between the bulk and the boundary, we consider the holographic dictionary $\mu_{CFT}=\mu_{bulk}$ and $C=N_{bulk}$.
The first law and the Smarr relation is given as:-
\begin{equation}
\label{eq:firstlaw}
\begin{split}
& dM=TdS+\tilde{\Phi}_ed\tilde{Q}_e+\tilde{\Phi}_md\tilde{Q}_m+\mu dC\\
& M=TS+\tilde{\Phi}_e\tilde{Q}_e+\tilde{\Phi}_m\tilde{Q}_m+\mu C
\end{split}
\end{equation}
Using \eqref{eq:firstlaw}, we get the Gibbs-Duhem relation:-
\begin{equation}
\begin{split}
   & d\mu=-\tilde{\mathcal{Q}}_e d\tilde{\Phi}_e-\tilde{\mathcal{Q}}_e d\tilde{\Phi}_e-\mathcal{S}dT\\ &\tilde{\mathcal{Q}}_e=\tilde{Q}_e/C, \quad \tilde{\mathcal{Q}}_m=\tilde{Q}_m/C, \quad \mathcal{S}
=S/C
\end{split}
\end{equation}
where $\tilde{\mathcal{Q}}_e$ is the electric charge per unit central charge, $\tilde{\mathcal{Q}}_e$ is the electric charge per unit central charge and $\mathcal{S}$ is entropy per unit $C-$ charge.

\section{Equation of states and homogeneity using the Kaniadakis statistics}
\label{sec:EOS}
In the Kaniadakis statistics, the entropy is given as:-
\begin{equation}
\label{eq:entropy}
    S=\frac{1}{\kappa}\sinh \left(\frac{\pi  \kappa  r^2}{G}\right)
\end{equation}
where $\kappa$ is the deformation parameter which quantifies the deformation from the standard Boltzmann-Gibbs statistical framework.
Solving the value of $r$ from \eqref{eq:entropy} and putting the rescaled charges and the value of $G$ from \eqref{eq:centralcharge} in \eqref{eq:mass1} we get:-
\begin{equation}
  M=\frac{\pi  C \kappa  \sinh ^{-1}(\kappa  S)+\pi ^2 \kappa ^2 \left(\tilde{Q}_e^2+\tilde{Q}_m^2\right)+\sinh ^{-1}(\kappa  S)^2}{2 \pi ^{3/2} C \kappa ^2 \sqrt{l^2 \sinh ^{-1}(\kappa  S)C \kappa^3}}  
\end{equation}
We series expand the mass for a small deformation parameter $\kappa$ and get:-
\begin{equation}
\label{eq:massf}
\small
    M=\frac{\pi  C S \left(12-\kappa ^2 S^2\right)+\pi ^2 \left(\tilde{Q}_e^2+\tilde{Q}_m^2\right) \left(\kappa ^2 S^2+12\right)-3 S^2 \left(\kappa ^2 S^2-4\right)}{24 \pi ^{3/2}  \sqrt{l^2 SC}}
\end{equation}
and the temperature $T$ and other variables are given as:-
\begin{equation}
\small
\label{eq:temperature}
    T=\frac{ \left(\pi  C S \left(12-5 \kappa ^2 S^2\right)+3 \pi ^2 \left(\tilde{Q}_e^2+\tilde{Q}_m^2\right) \left(\kappa ^2 S^2-4\right)-21 \kappa ^2 S^4+36 S^2\right)}{48 \pi ^{3/2} C^2 l \left(S/C\right)^{3/2}}
\end{equation}
\begin{equation}
\label{eq:electriccharge}
    \tilde{\Phi}_e=\frac{\sqrt{\pi} \tilde{Q}_e \left(\kappa ^2 S^2+12\right)}{12  \sqrt{l^2 SC}}, \quad  \tilde{\Phi}_m=\frac{\sqrt{\pi} \tilde{Q}_m \left(\kappa ^2 S^2+12\right)}{12  \sqrt{l^2 SC}}
\end{equation}
\begin{equation}
\label{eq:chemicalpotential}
\small
    \mu=\frac{\pi  C S\left(12-\kappa ^2 S^2\right)-\pi ^2 \left(\tilde{Q}_e^2+\tilde{Q}_m^2\right) \left(\kappa ^2 S^2+12\right)+3 S^2 \left(\kappa ^2 S^2-4\right)}{48 \pi ^{3/2}  \sqrt{l^2 S C^2}}
\end{equation}
If we rescale $S\rightarrow{}\lambda S$,  $\tilde{Q}_e\rightarrow{}\lambda \tilde{Q}_e$,  $\tilde{Q}_m\rightarrow{}\lambda\tilde{Q}_m$, $C\rightarrow{}\lambda C$, $\kappa\rightarrow{}\frac{1}{\lambda} \kappa$ and put it in the above equation, we see that the mass is rescaled as $M=\lambda M$ showing a first order homogeneity of mass M and the rest of the variables say $T=\lambda^0 T$ etc do not get rescaled like $M$ indicating a zeroth order homogeneity.

\subsection{Thermodynamic process and phase transitions}
The various thermodynamic processes are being studied for the dyonic AdS black holes in the RPST formalism using the Kaniadakis statistics. To find the turning points we use the equation:-
\begin{equation}
\label{eq:differential}
    \left(\frac{\partial T}{\partial S}\right)_{\tilde{Q}_e,\tilde{Q}_m,C}=0, \quad \left(\frac{\partial^2 T}{\partial S^2}\right)_{\tilde{Q}_e,\tilde{Q}_m,C}=0 
\end{equation}
Using this equation, solving the critical values of electric charge $\tilde{Q}_e$, entropy $S$ and temperature $T$ as:-
\begin{equation}
\label{eq:criticalvalues}
\begin{split}
&  \tilde{Q}_e^c=\frac{\sqrt{A_1^3}\sqrt{-\frac{\pi ^2 \kappa ^2 \tilde{Q}_m^2+12}{A_1}-\frac{12 \pi ^2 \tilde{Q}_m^2}{A_1^3}+35 \kappa ^2 A_1+\frac{4 \pi  C}{A_1^2}+5 \pi  C \kappa ^2}}{\pi  \sqrt{\kappa ^2 A_1^2+12}}\\
& S_c=A_1\\
& T_c=\frac{5 \pi  C \kappa ^4 A_1^4-3360 \kappa ^2 A_1^3-384 \pi  C \kappa ^2 A_1^2+1152 A_1+48 \pi  C}{60 \pi ^{3/2} \left(\kappa ^2 A_1^2+12\right) \sqrt{l^2 A_1C}}
\end{split}
\end{equation}
where $A_1$ is in Appendix. By introducing some relational variables $t=T/T_c$, $s=S/S_c$ and $q_e=\tilde{Q}_e/\tilde{Q}_e^c$ in equation \eqref{eq:temperature}, the EOS is:-
\begin{equation}
    t=\frac{t_1}{t_2}
\end{equation}where $t_1$ and $t_2$ are in Apendix.

\begin{figure}[htp]
    \centering
    \begin{subfigure}[b]{0.2\textwidth}
        \centering
        \includegraphics[width=\textwidth]{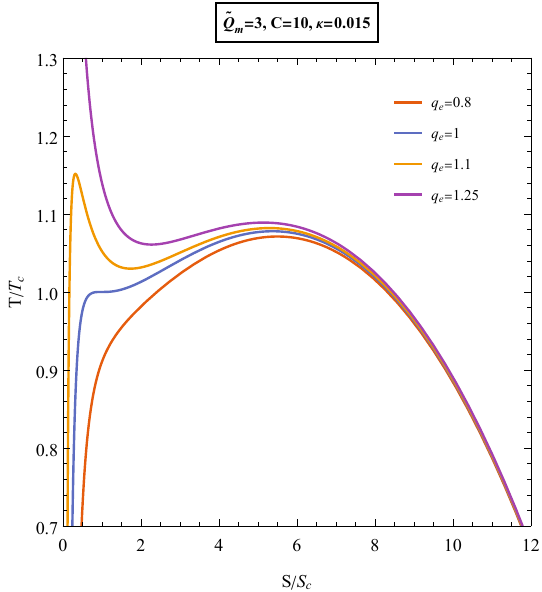} 
        \caption{}
        \label{fig:figure1}
    \end{subfigure}
    \begin{subfigure}[b]{0.2\textwidth}
        \centering
\includegraphics[width=\textwidth]{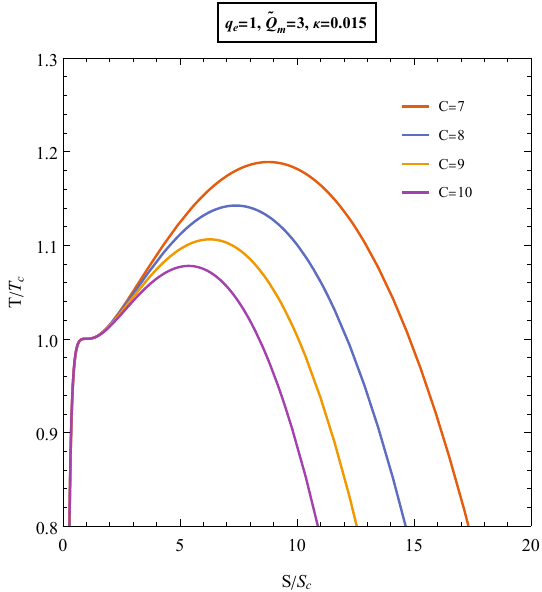} 
        \caption{}
        \label{fig:figure2}
    \end{subfigure}
    \begin{subfigure}[b]{0.2\textwidth}
        \centering
        \includegraphics[width=\textwidth]{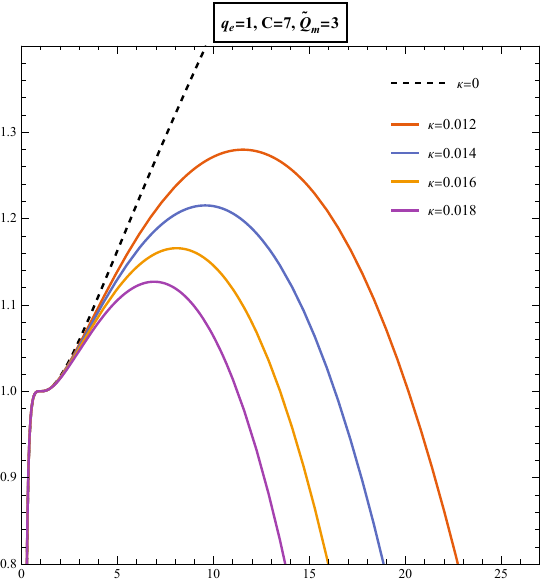} 
        \caption{}
        \label{fig:figure3}
    \end{subfigure}
    \begin{subfigure}[b]{0.2\textwidth}
        \centering
\includegraphics[width=\textwidth]{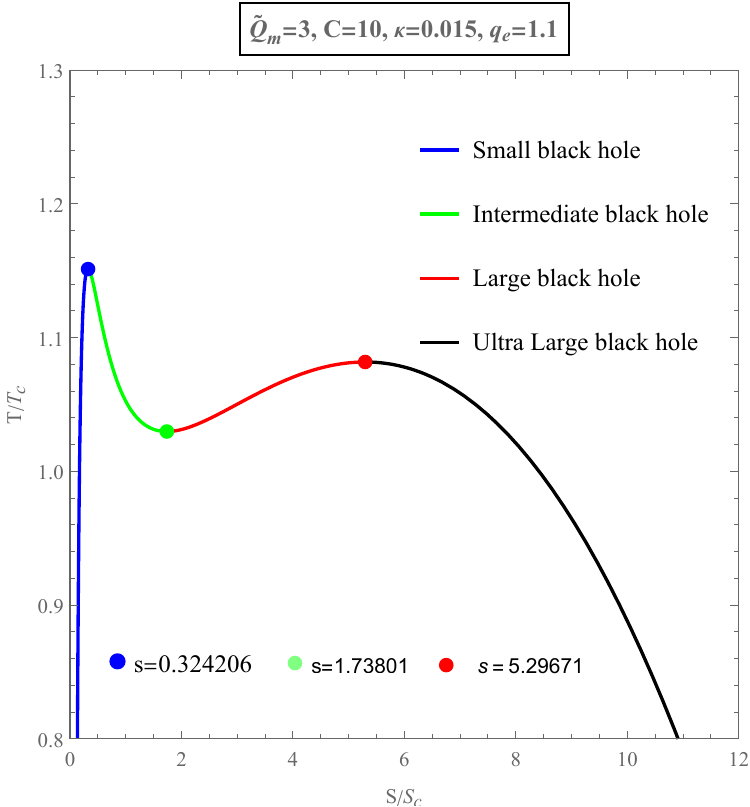} 
        \caption{}
        \label{fig:figure4}
    \end{subfigure}
    \caption{$T-S$ plots}
    \label{fig:one}
\end{figure}
The Helmholtz free energy is given using \eqref{eq:mass1}, \eqref{eq:temperature} as:-
\begin{equation}
\small
    F=\frac{3 \pi  C S \left(\kappa ^2 S^2+4\right)-\pi ^2 \left(\tilde{Q}_e^2+\tilde{Q}_m^2\right) \left(\kappa ^2 S^2-36\right)+3 S^2 \left(5 \kappa ^2 S^2-4\right)}{48 \pi ^{3/2}  \sqrt{l^2 S C}}
\end{equation}
By using the critical values given in \eqref{eq:criticalvalues} we get the critical values of free energy as:-
\begin{equation}
\small
    F_c=\frac{\left(-\pi C \kappa ^4 A_1^4+3360 \kappa ^2 A_1^3+480 \pi  C \kappa ^2 A_1^2-1152 A_1+528 \pi  C\right) \sqrt{ A_1}}{84 \pi ^{3/2} l \sqrt{C} \left(\kappa ^2 A_1^2+12\right)}
\end{equation}
By introducing $f=F/F_c$ we can write:-
\begin{equation}
    f=-\frac{f_1}{f_2}
\end{equation}
where $f_1$ and $f_2$ are in Appendix.
\begin{figure}[ht]
    \centering
    \begin{subfigure}[b]{0.2\textwidth}
        \centering
        \includegraphics[width=\textwidth]{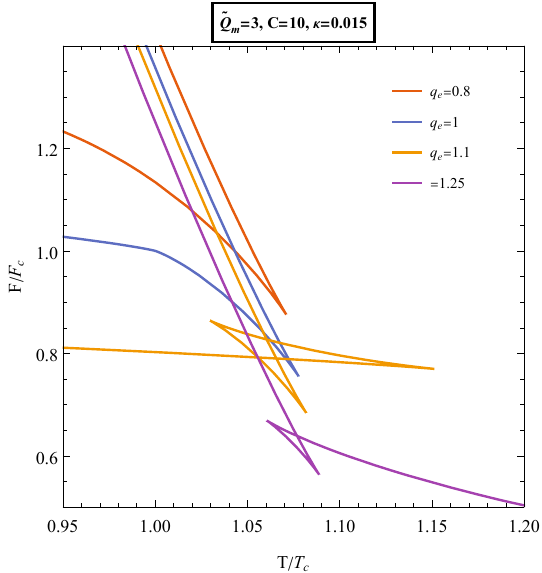} 
        \caption{}
        \label{fig:figure5}
    \end{subfigure}
    \begin{subfigure}[b]{0.2\textwidth}
        \centering
        \includegraphics[width=\textwidth]{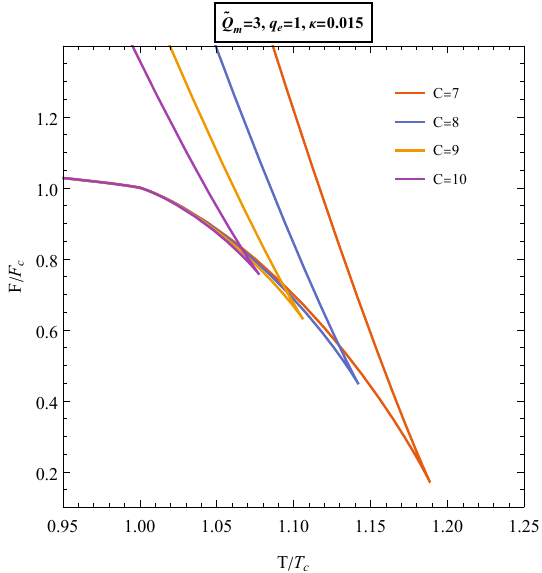} 
        \caption{}
        \label{fig:figure6}
    \end{subfigure}
    \begin{subfigure}[b]{0.2\textwidth}
        \centering
        \includegraphics[width=\textwidth]{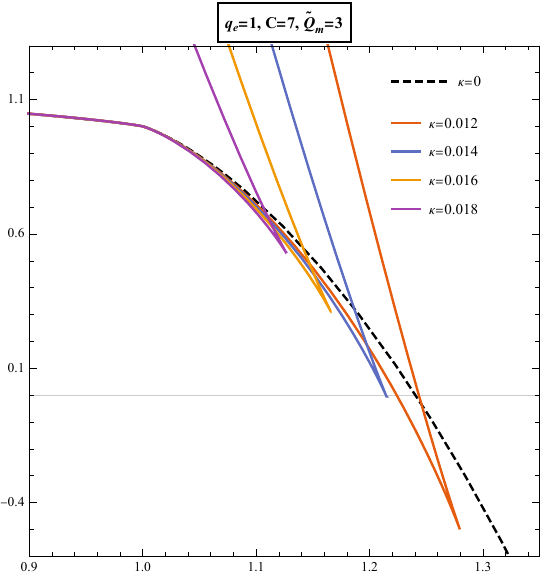} 
        \caption{}
        \label{fig:figure7}
    \end{subfigure}
    \begin{subfigure}[b]{0.2\textwidth}
        \centering
        \includegraphics[width=\textwidth]{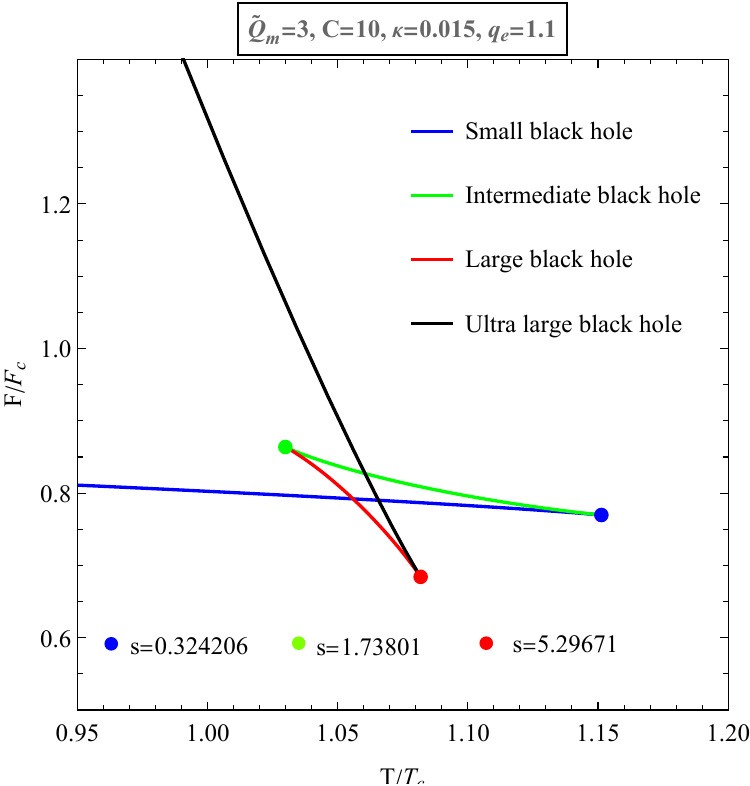} 
        \caption{}
        \label{fig:figure8}
    \end{subfigure}
    \caption{$F-T$ plots}
    \label{fig:two}
\end{figure}
The iso-$e$- charge processes are shown in the $T-S$ and $F-T$ plots in Fig \ref{fig:one} and \ref{fig:two}. In fig \ref{fig:figure1} and \ref{fig:figure5} we see a van der Waals type phase transition of order one (yellow) for various supercritical values of electric charge $\tilde{Q}_e$. At the critical value (blue), we see a second-order phase transition and in the $F-T$ plot it is seen as a kink, and for the subcritical values (red), we see a monotonous plot up to a certain entropy point. In the purple plot, we see a non-equilibrium phase transition which corresponds to the Hawking-Page phase transition, caused due to the inclusion of the magnetic charge $\tilde{Q}_m$. One thing that separates from the other entropy models is the inclusion of an extra branch at higher entropy which we have labeled as ultra-large black holes. We can see in figure \ref{fig:figure4} and \ref{fig:figure8}. By setting the parameters for a first-order phase transition, we see four branches namely the small-intermediate-large-ultra large black holes each color-coded for better visuality, and also the points where the branch's ends are shown. In figure \ref{fig:figure2} and \ref{fig:figure6}, we plot for various values of central charge $C$ and note that as the $C$ decreases the ultra-large black hole branch becomes bigger for a larger entropy. It also signals that the $C$ in a way controls the deformation or inclusion of the ultra-large black holes. Plotting for various values of the deformation parameter $\kappa$ in fig. \ref{fig:figure3}, \ref{fig:figure7}, when $\kappa=0$ it goes back to the classical Boltzmann-Gibbs statistics, but as the $\kappa$ increases which quantifies the deviation for the classical Boltzmann-Gibbs statistics, we see the deformation is more pronounced and also looks similar to the plots with a variation of $C$. All the plots are plots at the critical electric charge $\tilde{Q}_m$.\\
We continue the study by setting $\tilde{Q}_m=0$. With this condition, the $T-S$ and $F-T$ processes appear different by losing the non-equilibrium transition and Hawking-Page type phase transition. Under this condition, the temperature is:
\begin{equation}
\label{eq:TempQ}
    T=\frac{\left(\pi  C S \left(12-5 \kappa ^2 S^2\right)+3 \pi ^2 \tilde{Q}_e^2 \left(\kappa ^2 S^2-4\right)-21 \kappa ^2 S^4+36 S^2\right)}{48 \pi ^{3/2} C^2 l \left(\frac{S}{C}\right)^{3/2}}
\end{equation}
Using equation \eqref{eq:differential}, the critical values of entropy and charge and temperature is:-
\begin{equation}
\begin{split}
\small
\label{eq:crticalvalues1}
&    S_c=A_1\\
&   \tilde{Q}_e^c= \frac{\sqrt{35 \kappa ^2 A_1^4-12 A_1^4+4 \pi  C A_1+5 \pi  C A_1^3 \kappa ^2}}{\pi  \sqrt{\kappa ^2 A_1^2+12}}\\
& T_c=\frac{ \left(-21 \kappa ^2 A_1^4+36 A_1^2+\pi  C \left(12-5 \kappa ^2 A_1^2\right) A_1+D_3\right)}{48 \pi ^{3/2} C^2 l \left(\frac{ A_1}{C}\right)^{3/2}}
\end{split}
\end{equation}
where $A_1$ and $D_3$ are in Appendix.
By introducing the relational variables studied before in equation \eqref{eq:TempQ}, the EOS is:-
\begin{equation}
    t=\frac{D_4}{D_5}
\end{equation}
where $D_4$ and $D_5$ are in Appendix.
\begin{figure}[htp]
    \centering
    \begin{subfigure}[b]{0.2\textwidth}
        \centering
        \includegraphics[width=\textwidth]{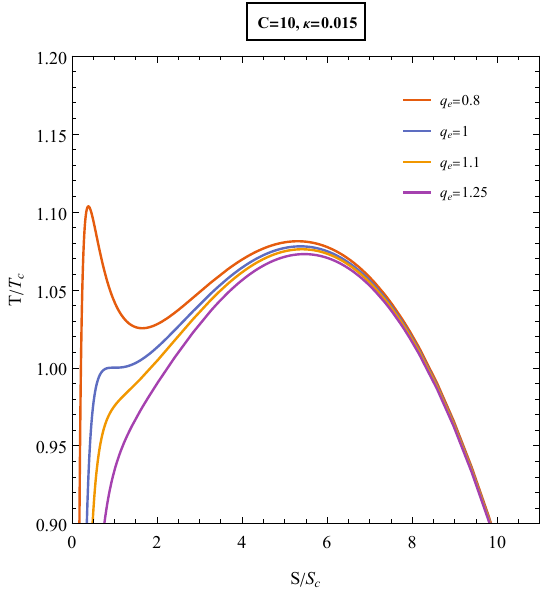} 
        \caption{}
        \label{fig:figure9}
    \end{subfigure}
    \begin{subfigure}[b]{0.2\textwidth}
        \centering
        \includegraphics[width=\textwidth]{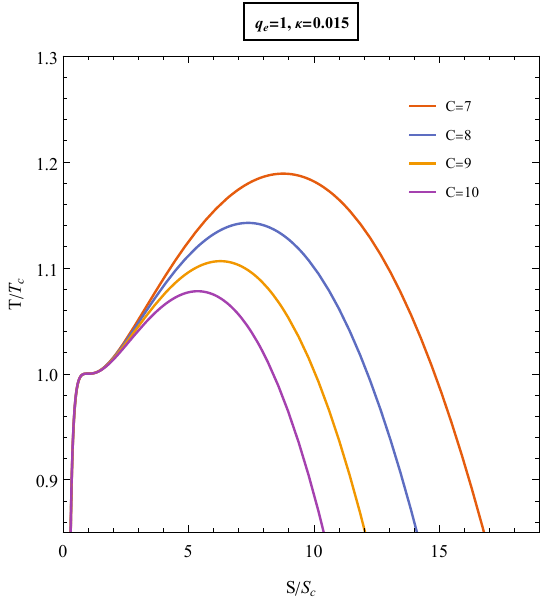} 
        \caption{}
        \label{fig:figure10}
    \end{subfigure}
    \begin{subfigure}[b]{0.2\textwidth}
        \centering
        \includegraphics[width=\textwidth]{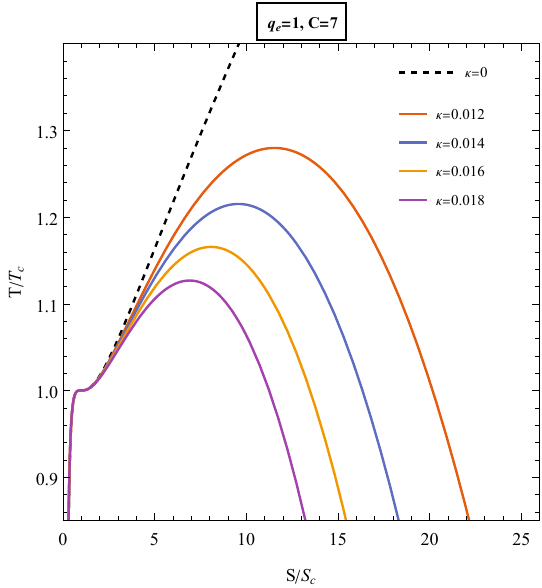} 
        \caption{}
        \label{fig:figure11}
    \end{subfigure}
    \caption{$T-S$ plots}
    \label{fig:three}
\end{figure}

The free energy is given as:-
\begin{equation}
    F=\frac{3 \pi  C S \left(\kappa ^2 S^2+4\right)+\pi ^2 \tilde{Q}_e^2 \left(36-\kappa ^2 S^2\right)+3 S^2 \left(5 \kappa ^2 S^2-4\right)}{48 \pi ^{3/2}  \sqrt{l^2 S C}}
\end{equation}
By using all the critical values and introducing $f=F/F_c$, free energy equation is written as:-
\begin{equation}
\small
    f=-\frac{D_6}{4 \left(\pi  C \kappa ^4 A_1^4-3360 \kappa ^2 A_1^3-480 \pi  C \kappa ^2 A_1^2+1152 A_1-528 \pi  c\right) \sqrt{s}}
\end{equation}
where using critical values, $F_C$ is given as:-
\begin{equation}
\small
   F_c=\frac{\left(-\pi C \kappa ^4 A_1^4+3360 \kappa ^2 A_1^3+480 \pi  C \kappa ^2 A_1^2-1152 A_1+528 \pi  C\right) \sqrt{A_1}}{84 \pi ^{3/2} l \sqrt{C} \left(\kappa ^2 A_1^2+12\right)} 
\end{equation}

\begin{figure}[h]
    \centering
    \begin{subfigure}[b]{0.2\textwidth}
        \centering
        \includegraphics[width=\textwidth]{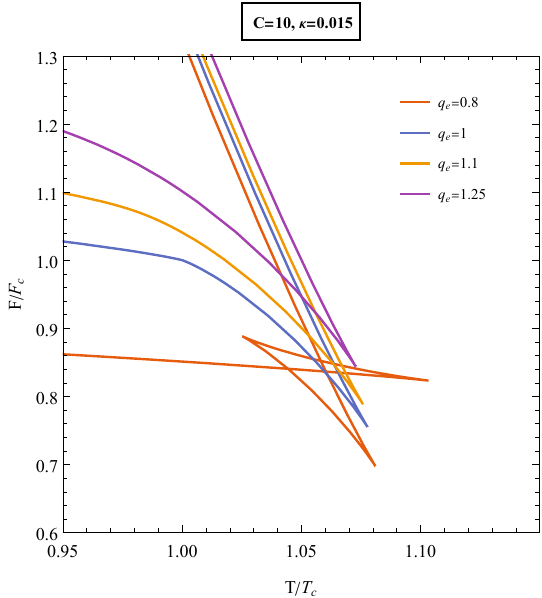} 
        \caption{}
        \label{fig:figure12}
    \end{subfigure}
    \begin{subfigure}[b]{0.2\textwidth}
        \centering
        \includegraphics[width=\textwidth]{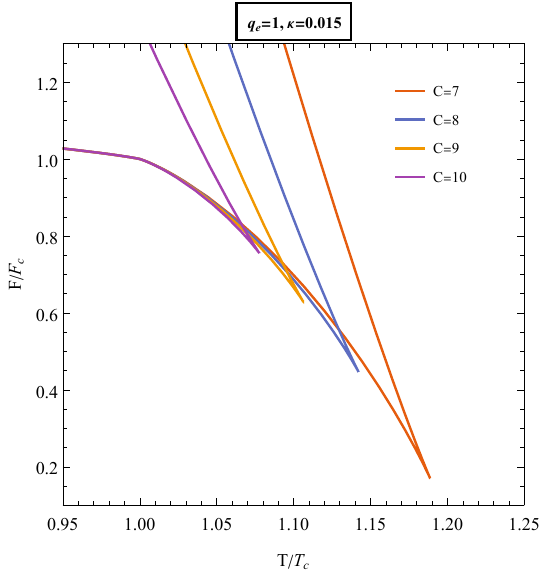} 
        \caption{}
        \label{fig:figure13}
    \end{subfigure}
    \begin{subfigure}[b]{0.2\textwidth}
        \centering
        \includegraphics[width=\textwidth]{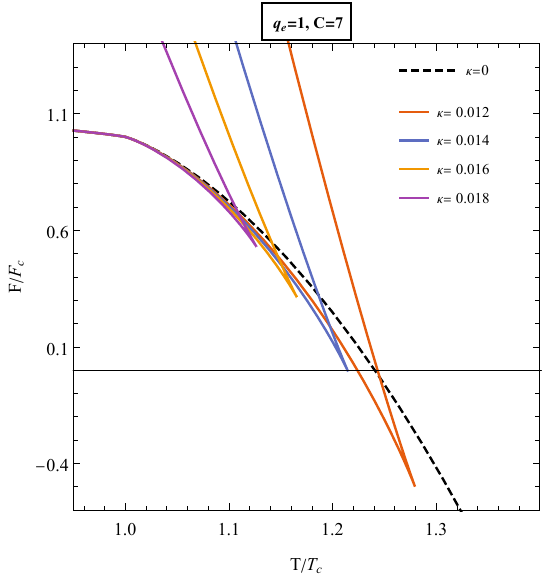} 
        \caption{}
        \label{fig:figure14}
    \end{subfigure}
    \caption{$F-T$ plots}
    \label{fig:four}
\end{figure}
 We see that in Fig. \ref{fig:three} and \ref{fig:four} due to the absence of the magnetic charge, the non-equilibrium transition in the $T-S$ plot and the Hawking-Page phase transition is absent rest all are the same matching Fig \ref{fig:one} and \ref{fig:two}. But one thing surprising is that now for subcritical values of the critical electric charge, we see Van der Waals type phase transition of order one in figure \ref{fig:figure9} which was seen for supercritical values in fig \ref{fig:figure1}. One can see that as the $\tilde{Q}_m$ turns off it affects the way the transition happened or the order of the transition has interchanged.\\
Setting both the electric $\tilde{Q}_e$ and magnetic $\tilde{Q}_m$ charge to zero, the system reduces to a Schwarzschild AdS black hole. But due to the $\kappa$ parameter, it does quite reduce to that. The temperature and the free energy is written as:-
\begin{equation}
\begin{split}
\label{eq:tempfree00}
  &  T=\frac{ \left(\pi  C S \left(12-5 \kappa ^2 S^2\right)-21 \kappa ^2 S^4+36 S^2\right)}{48 \pi ^{3/2} C^{1/2} l S^{3/2}}\\
  &F=\frac{\sqrt{ S} \left(\pi  C \left(\kappa ^2 S^2+4\right)+S \left(5 \kappa ^2 S^2-4\right)\right)}{16 \pi ^{3/2} l \sqrt{C}}
  \end{split}
\end{equation}
using \eqref{eq:differential}, the critical values of entropy, temperature, and free energy are:-
\begin{equation}
    S_c=\frac{1}{525} B_1
\end{equation}
\begin{equation}
    T_c=\frac{280.61 \left(\frac{1}{525} C \pi  \left(12-\frac{\kappa ^2 B_1^2}{55125}\right) B_1+\frac{4 B_1)^2}{30625}-\frac{\kappa ^2B_1^4}{3617578125}\right)}{ C^{1/2} \pi ^{3/2} lB_1^{3/2}}
\end{equation}
\begin{equation}
    F_c=\frac{\left(\frac{1}{525} B_1 \left(\frac{\kappa ^2 B_1^2}{55125}-4\right)+\pi  C \left(\frac{\kappa ^2B_1^2}{275625}+4\right)\right) \sqrt{ B_1}}{80 \sqrt{21} \pi ^{3/2} l\sqrt{C}}
\end{equation}
Where $B_1$ is in Appendix. Using the relative parameters $t=T/T_c$, $f=F/F_c$ and $s=S/S_c$ the EOS and the free energy is written as:-
\begin{equation}
\label{eq:tsk001}
   t=\frac{ \left(\frac{1}{525} C \pi  s \left(12-\frac{s^2 \kappa ^2 B_1^2}{55125}\right)B_1+\frac{4s^2 B_1^2}{30625}-\frac{s^4 \kappa ^2 B_1^4}{3617578125}\right)}{s^{3/2} \left(\frac{1}{525} C\pi  \left(12-\frac{\kappa ^2B_1^2}{55125}\right) B_1+\frac{4 B_1^2}{30625}-\frac{\kappa ^2 B_1^4}{3617578125}\right)} 
\end{equation}
\begin{equation}
\label{eq:ftk001}
    f=\frac{\sqrt{s} \left(\frac{1}{525} s B_1 \left(\frac{s^2 \kappa ^2 B_1^2}{55125}-4\right)+C \pi  \left(\frac{s^2 \kappa ^2 B_1^2}{275625}+4\right)\right)}{ \left(\frac{1}{525} B_1 \left(\frac{\kappa ^2B_1^2}{55125}-4\right)+C \pi  \left(\frac{\kappa ^2 B_1^2}{275625}+4\right)\right)}
\end{equation}

\begin{figure}[htbp]
    \centering
    \begin{subfigure}[b]{0.2\textwidth}
        \centering
        \includegraphics[width=\textwidth]{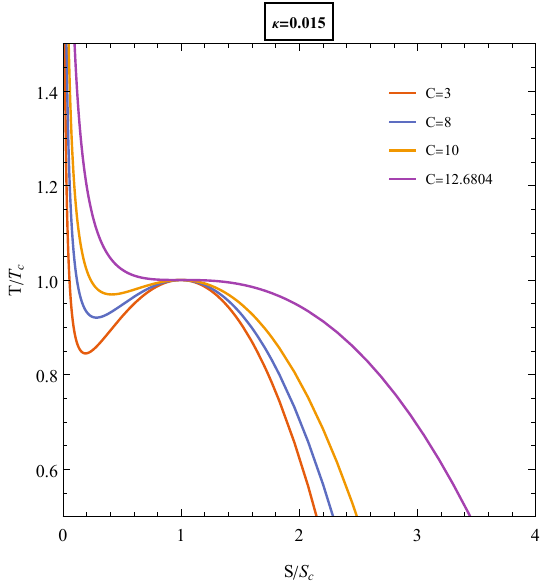} 
        \caption{}
        \label{fig:figure15}
    \end{subfigure}
    \begin{subfigure}[b]{0.2\textwidth}
        \centering
        \includegraphics[width=\textwidth]{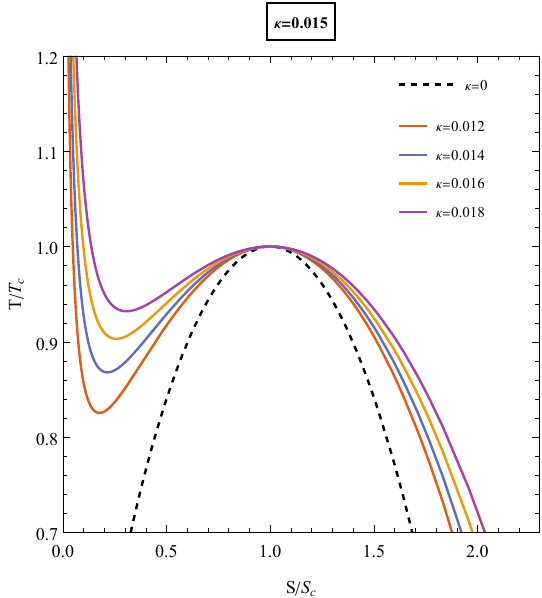} 
        \caption{}
        \label{fig:figure16}
    \end{subfigure}
    \caption{$T-S$ plots}
    \label{fig:five}
\end{figure}

\begin{figure}[htp]
    \centering
    \begin{subfigure}[b]{0.2\textwidth}
        \centering
        \includegraphics[width=\textwidth]{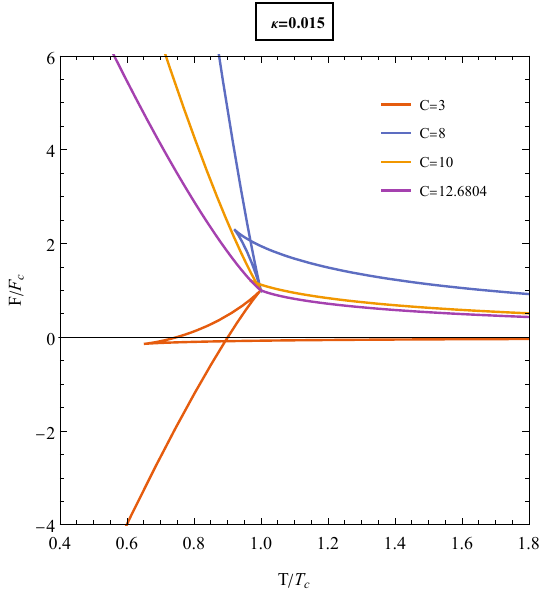} 
        \caption{}
        \label{fig:figure17}
    \end{subfigure}
    \begin{subfigure}[b]{0.2\textwidth}
        \centering
        \includegraphics[width=\textwidth]{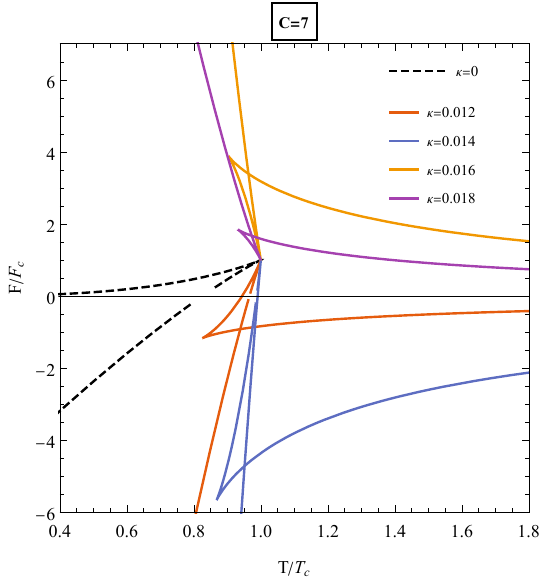} 
        \caption{}
        \label{fig:figure18}
    \end{subfigure}
    \caption{$F-T$ plots}
    \label{fig:six}
\end{figure}
Using the equation \eqref{eq:tsk001} and \eqref{eq:ftk001} we plot the $T-S$ and $F-T$ plots seen in Fig. \ref{fig:five} and \ref{fig:six}. In fig \ref{fig:figure15}, \ref{fig:figure17} we plot for the various values of $C$. We see three branches corresponding to the small (unstable), large (stable), ultra-large (unstable) black holes, but at a critical value of $C=12.6804$, we observe only two branches for small (unstable),  ultra-large (unstable) black holes. One thing to notice in the $F-T$ plot is that in the ultra-large black hole branch the free energy shifts from a negative value to a positive value. In figure \ref{fig:figure18} and \ref{fig:figure18} we plot for the various values of $\kappa$ and we see the same kind of transition as we plot for the $C$, but when $\kappa\approx0$ we see only two branches namely the large(stable) and the ultra (unstable) black holes.
\subsection{Double-superfluid $\lambda$ phase transition}
 We also see the $T-S$ processes for the fixed $\tilde{\Phi}_e$ by differentiating the mass w.r.t charge $\tilde{Q}_e$ getting $\tilde{\Phi}_e$ and hence writing the temperature $T(\tilde{Q}_m,\tilde{\Phi}_e,C,S)$ as:-
 \begin{equation}
 \scriptsize
     T=\frac{\left(3 \pi ^2 \left(\kappa ^2S^2-4\right) \left(\frac{144 C l^2 S \tilde{\Phi}_e^2}{\pi  \left(\kappa ^2 S^2+12\right)^2}+\tilde{Q}_m^2\right)+\pi  C S \left(12-5 \kappa ^2 S^2\right)-21 \kappa ^2 S^4+36 S^2\right)}{48 \pi ^{3/2}l C^{1/2} S^{3/2}}
 \end{equation}
Using \eqref{eq:differential}, the critical points are:-
\begin{equation}
\small
  \tilde{\Phi}_e^c= \frac{\sqrt{A_4^2 \left(\kappa ^2 A_4^2+12\right)^3}\sqrt{5 C \pi  \kappa ^2+35 \kappa ^2 A_4-\frac{\pi ^2 \tilde{Q}_m^2 \kappa ^2+12}{A_4}+\frac{4 C \pi }{A_4^2}-\frac{12 \pi ^2 \tilde{Q}_m^2}{A_4^3}}}{12l \sqrt{\pi } \sqrt{C}}
\end{equation}
\begin{equation}
    S=A_5
\end{equation}
\begin{equation}
    T_c=\frac{D_7}{24 C^{1/2} \pi ^{3/2} \left(l^2 A_5\right)^{3/2} \left(5 \kappa ^4 A_4^4-72 \kappa ^2 A_4^2-48\right)}
\end{equation}
Using the critical values given above, the EOS is:-
\begin{equation}
    t=\frac{ D_8}{D_9}
\end{equation}
where $D_8$ and $D_9$ are in Appendix.
\begin{figure}[htp]
    \centering
    \begin{subfigure}[b]{0.2\textwidth}
        \centering
        \includegraphics[width=\textwidth]{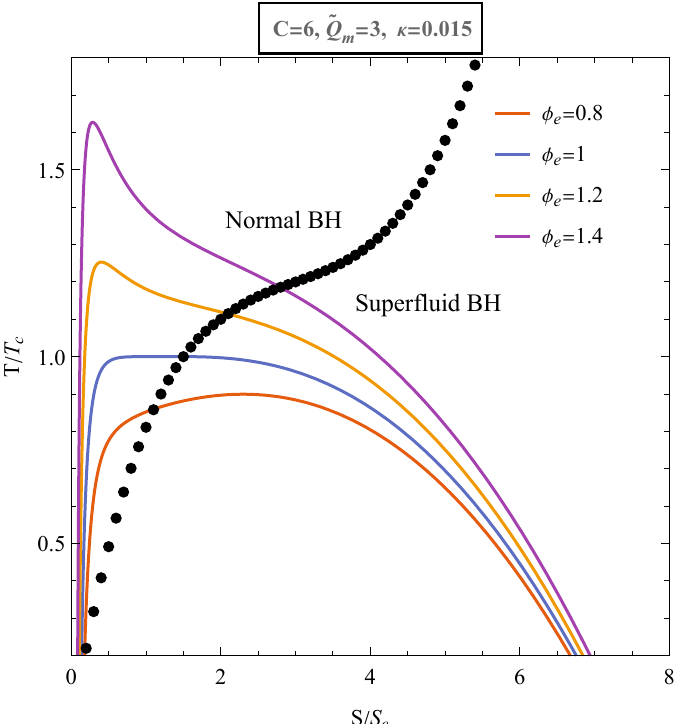} 
        \caption{$T-S$ plots with various values of $\tilde{\phi}_e$ and the dashed line are the critical points.}
        \label{fig:figure19}
    \end{subfigure}
    \begin{subfigure}[b]{0.2\textwidth}
        \centering
        \includegraphics[width=\textwidth]{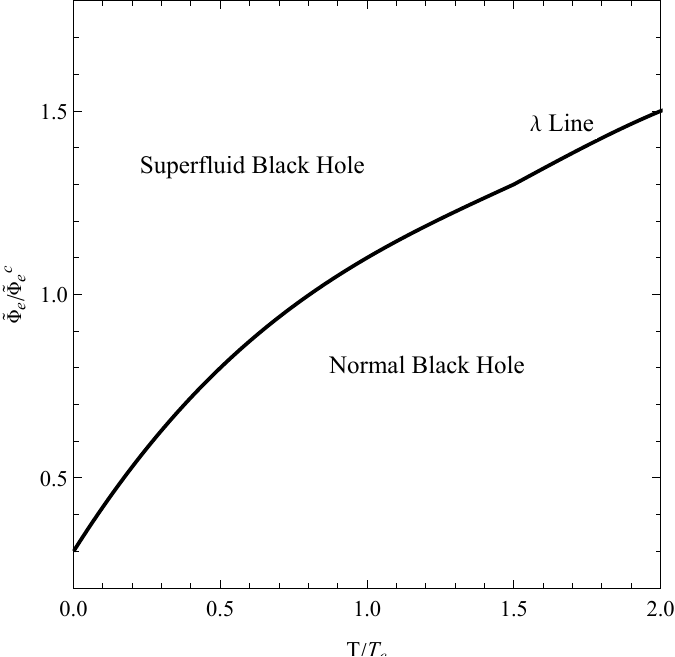} 
        \caption{Coexistence plot between $\tilde{\Phi}_e-T$ where the black dashed line is the line of critical points ($\lambda$ line).}
        \label{fig:figurelambda}
    \end{subfigure}
    \begin{subfigure}[b]{0.2\textwidth}
        \centering
        \includegraphics[width=\textwidth]{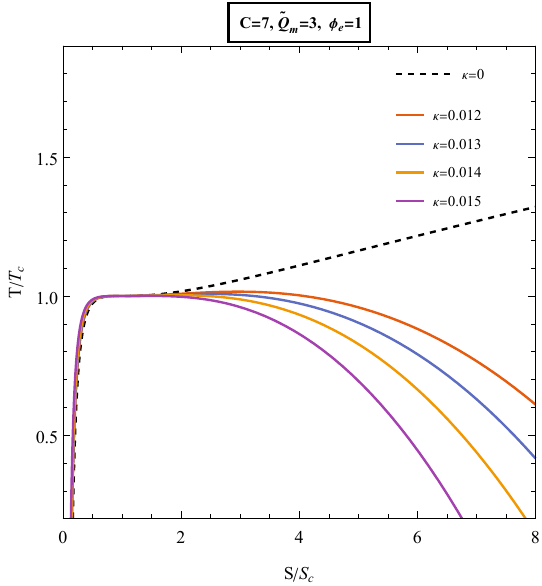} 
        \caption{$T-S$ plot for various values of deformation parameter $\kappa$.}
        \label{fig:figure20}
    \end{subfigure}
    \begin{subfigure}[b]{0.2\textwidth}
        \centering
        \includegraphics[width=\textwidth]{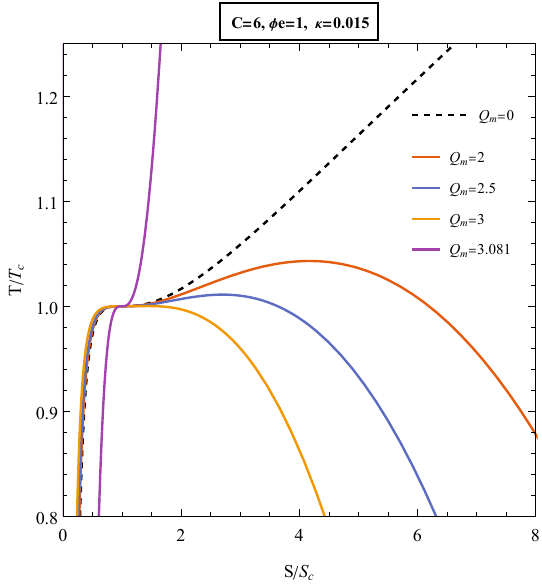} 
        \caption{$T-S$ plot for various values of $\tilde{Q}_m$.}
        \label{fig:figure21}
    \end{subfigure}
    \caption{}
    \label{fig:seven}
\end{figure}
For clear visual, we plot $\mu-T$ where $\mu$ is the free energy divided by the $C$. Putting $\tilde{Q}_e= \frac{12 \sqrt{C} \tilde{\Phi}_e \sqrt{l^2 S}}{\sqrt{\pi } \left(\kappa ^2 S^2+12\right)}$ in \eqref{eq:chemicalpotential} we obtain:-
\begin{equation}
\scriptsize
\mu=\frac{-\pi  C S\left(144 \left(l^2 \tilde{\Phi}_e^2-1\right)+\kappa ^4 S^4\right)-\pi ^2 \tilde{Q}_m^2 \left(\kappa ^2 S^2+12\right)^2+3 S^2 \left(\kappa ^4 S^4+8 \kappa ^2 S^2-48\right)}{48 \pi ^{3/2} c^{3/2} \left(\kappa ^2 S^2+12\right) \sqrt{l^2 S}}
\end{equation}
From the critical points, the critical value of $\mu$ is:-
\begin{equation}
    \mu_c=\frac{D_{10}}{24 C^{3/2} \pi ^{3/2} \sqrt{l^2 A_4} \left(5 \kappa ^4 A_4^4-72 \kappa ^2 A_4^2-48\right)}
\end{equation}
where $D_{10}$ is in Appendix.
Using the relative parameters $\tilde{m}=\mu/\mu_c$, $s=S/S_c$ and $\phi_e=\tilde{\Phi}_e/\tilde{\Phi}_e^c$ we get:-
\begin{equation}
        \tilde{m}=\frac{D_{11}}{D_{12}}
\end{equation}
\begin{figure}[htp]
    \centering
    \begin{subfigure}[b]{0.2\textwidth}
        \centering
        \includegraphics[width=\textwidth]{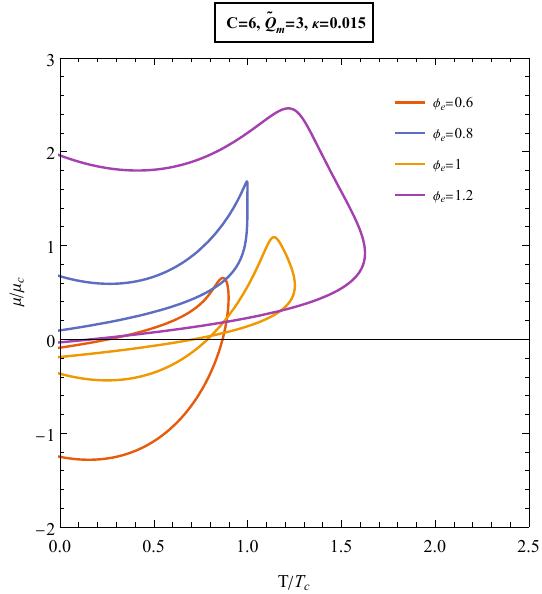} 
        \caption{}
        \label{fig:figure22}
    \end{subfigure}
    \begin{subfigure}[b]{0.2\textwidth}
        \centering
        \includegraphics[width=\textwidth]{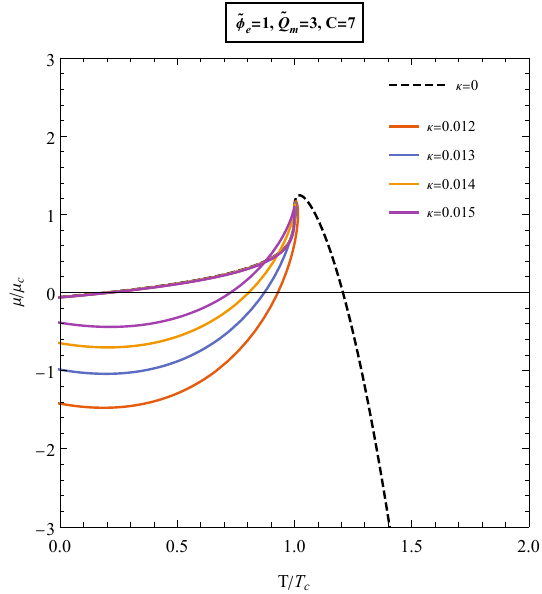} 
        \caption{}
        \label{fig:figure23}
    \end{subfigure}
    \begin{subfigure}[b]{0.2\textwidth}
        \centering
        \includegraphics[width=\textwidth]{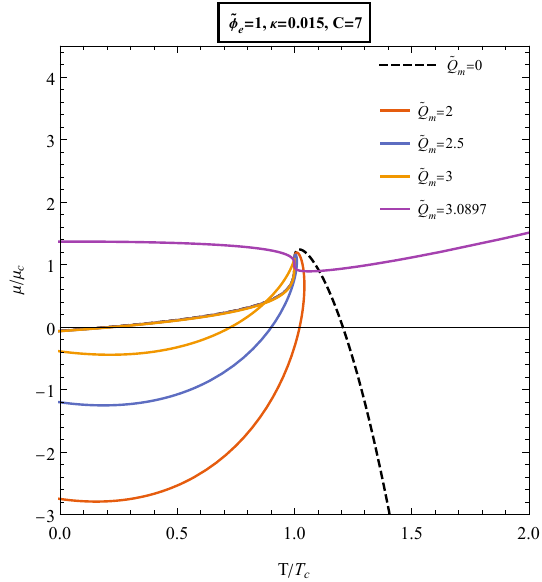} 
        \caption{}
        \label{fig:figure24}
    \end{subfigure}
    \caption{$T-S$ plots}
    \label{fig:eight}
\end{figure}
In Fig \ref{fig:seven} and \ref{fig:eight}, we plot $T-S$ and $\mu-T$ for various values of parameters. In fig \ref{fig:figure19} and \ref{fig:figure22}, we plot for various values of $\phi_e=\tilde{\Phi}_e/\tilde{\Phi}_e^c$ noticing only two types of branches namely small (stable) and ultra-large black hole branch (unstable). Visually it is seen that for supercritical values of $\phi_e$ we may obtain three branches of black holes but upon calculations, it is seen that for the purple plot in fig \ref{fig:figure19} and \ref{fig:figure22}, we see only one turning point at $s=0.286976$ depicting only two branches. At the critical point, we also obtain a novel double-superfluid $\lambda$ phase transition of order two which is quite exotic and different seen in the charged AdS black hole. In fig. \ref{fig:figure19} the line of the critical points is described by the dashed line. Within this line, we see the (continuous) second order $\lambda$ phase transition. This occurrence matches condensed matter systems such as the fluid/superfluid transitions, superconductivity, and paramagnetic/ferromagnetism transitions. Normally the dashed line as the entropy $S$ increases, the temperature $T$ should get constant but here we see it increases to infinite values. This is seen due to the inclusion of the deformation parameter $\kappa$. In fig.\ref{fig:figurelambda}, we see the coexistence plot where the $\lambda$ line shows the boundary of the coexistence region. Above the $\lambda$ line, the superfluid black hole dominates, and below the line, the normal black hole exists exclusively. In the $\lambda$ line both the black holes can exist simultaneously hence called the coexists line. In fig \ref{fig:figure20}, \ref{fig:figure23} we plot the various values of the deformation parameter $\kappa$. Here also we get two branches for small (stable) and ultra-large (unstable) black holes and the impact of $\kappa$ can be seen in the plots. As $\kappa$ reduces, the temperature $T/T_c \rightarrow{\infty}$ as the entropy $S/S_c \rightarrow{\infty}$. In the $F-T$ plot for $\kappa\approx0$, we get an infinite value of the free energy and as the value of $\kappa$ increases, we obtain a finite value of free energy. Now for the critical value of $\phi_e$ where we see the double superfluid $\lambda$ phase transition we plot for the various values of $\tilde{Q}_m$ where we obtain two branches, one small (stable) and other ultra-large (unstable) black hole branch where the turning point depends on the value of $\tilde{Q}_m$. At a particular value of $\tilde{Q}_m=3.081$, we only get one stable branch. It should be noted that there is a point of discontinuity at $s=1$ which can be seen clearly in the specific heat plot. In fig. \ref{fig:eight1}, we meticulously see the specific heat plots both for specific heat vs. entropy in figure \ref{fig:figure221}, \ref{fig:figure231}, \ref{fig:figure261} and specific heat vs. temperature in figure \ref{fig:figure241}, \ref{fig:figure251}. In \ref{fig:figure221} and \ref{fig:figure241}, we plot for a particular value of the variables and see various branches that go stable-stable-unstable from left to right seen in the figure. The discontinuous line that goes up between the large and ultra-large black hole meets at a point seen in figure \ref{fig:figure261}. It looks like the $\lambda$ structure is seen for the charged AdS black hole but there is a vertical line at the end due to the deformation parameter $\kappa$. At the critical points for $\tilde{Q}_m$, we see clearly $\lambda$ like structure in figure \ref{fig:figure231}, \ref{fig:figure251}.\\
\begin{figure}[htp]
    \centering
    \begin{subfigure}[b]{0.2\textwidth}
        \centering
        \includegraphics[width=\textwidth]{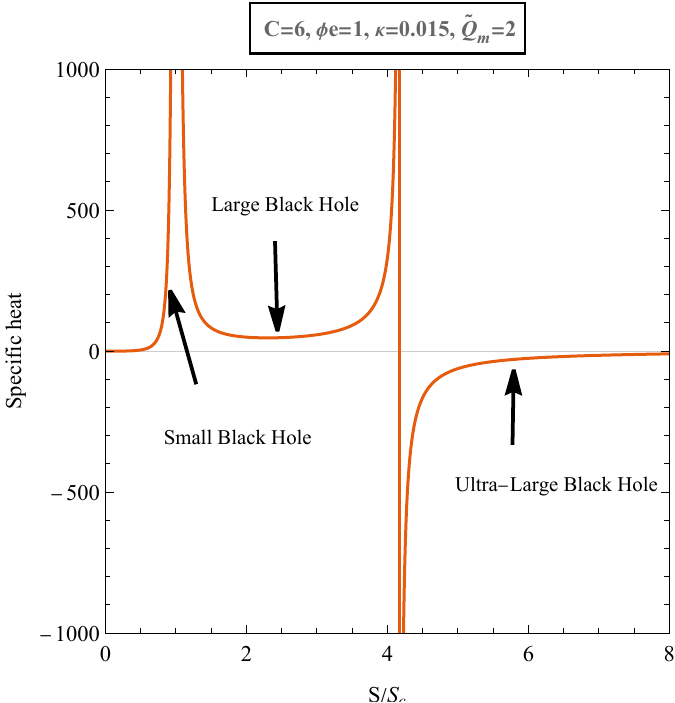} 
        \caption{}
        \label{fig:figure221}
    \end{subfigure}
    \begin{subfigure}[b]{0.2\textwidth}
        \centering
        \includegraphics[width=\textwidth]{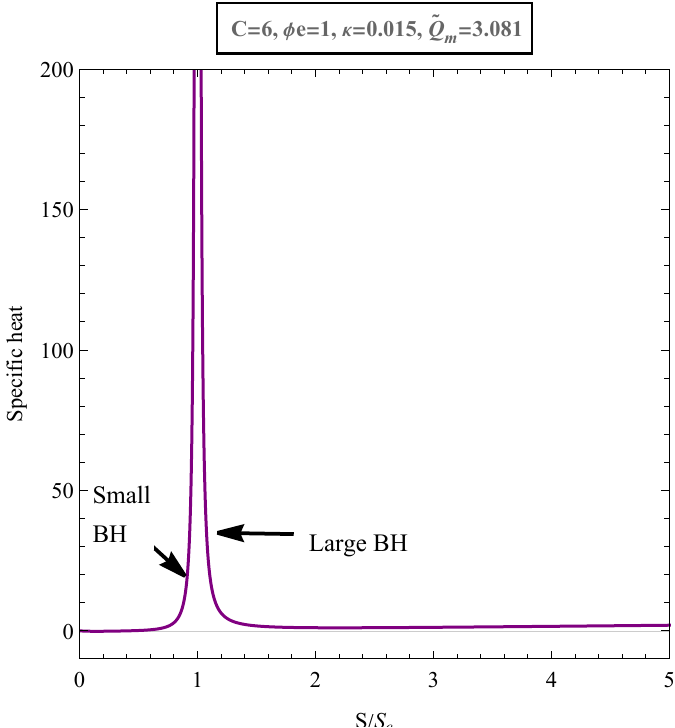} 
        \caption{}
        \label{fig:figure231}
    \end{subfigure}
    \begin{subfigure}[b]{0.2\textwidth}
        \centering
        \includegraphics[width=\textwidth]{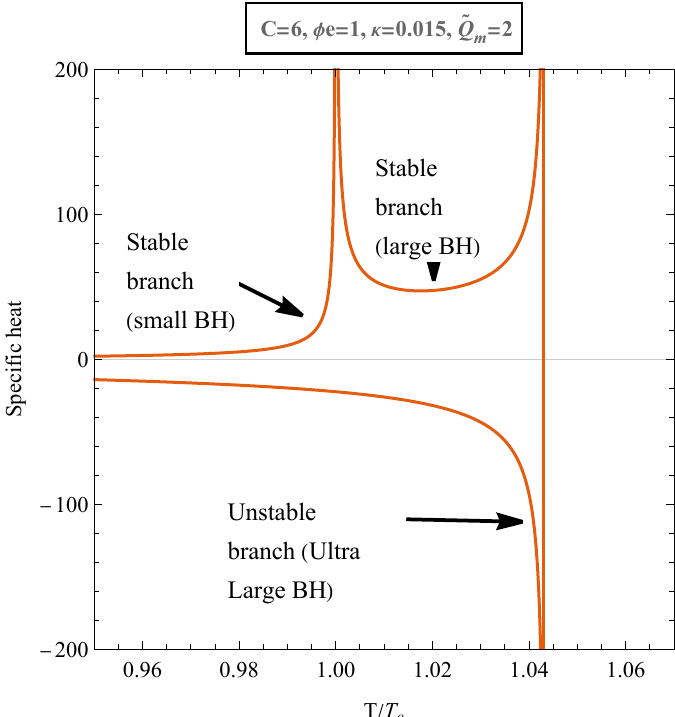} 
        \caption{}
        \label{fig:figure241}
    \end{subfigure}
     \begin{subfigure}[b]{0.2\textwidth}
        \centering
        \includegraphics[width=\textwidth]{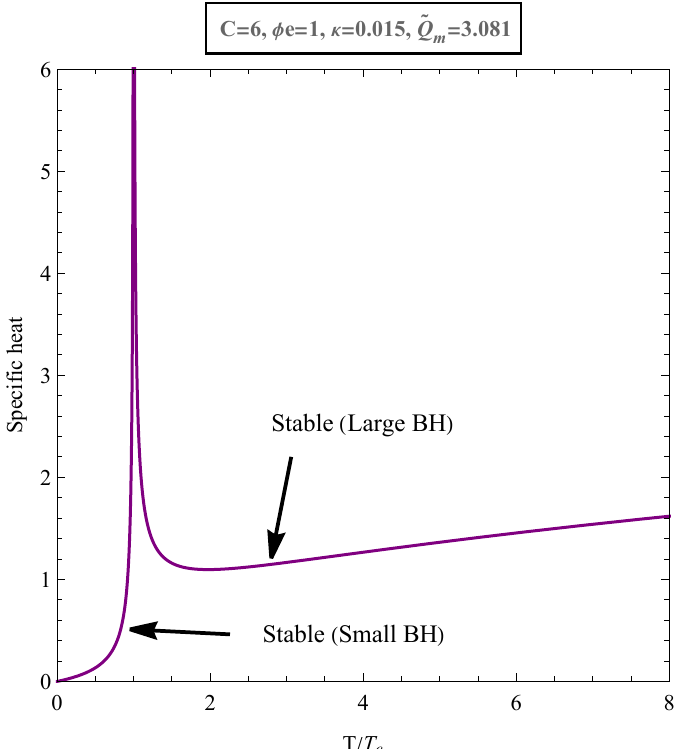} 
        \caption{}
        \label{fig:figure251}
    \end{subfigure}
     \begin{subfigure}[b]{0.2\textwidth}
        \centering
        \includegraphics[width=\textwidth]{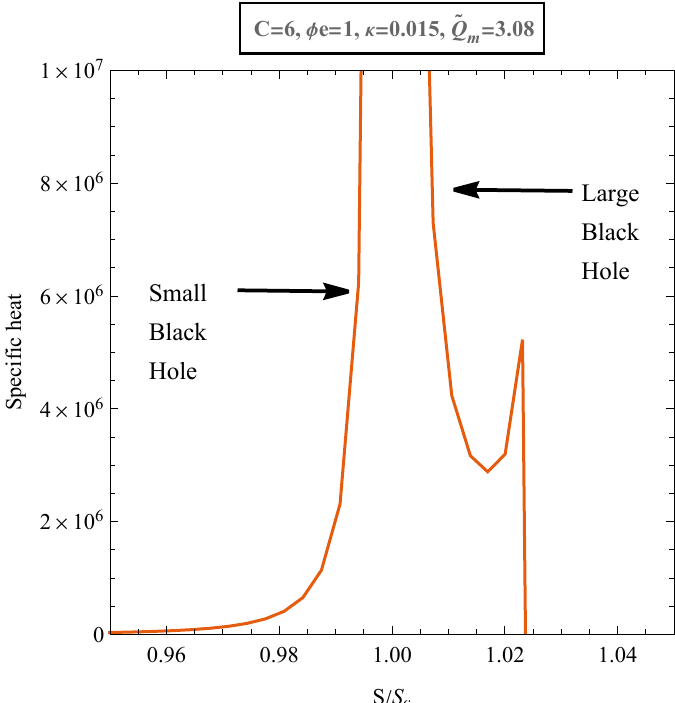} 
        \caption{}
        \label{fig:figure261}
    \end{subfigure}
    \caption{}
    \label{fig:eight1}
\end{figure}
Fixing the $\tilde{\Phi}_e$, $\tilde{\Phi}_m$ one can also study the $T-S$ and $\mu-T$ processes. The temperature and free energy is written as:-
\begin{equation}
\small
    T=\frac{\frac{432 \pi  C l^2 \left(\kappa ^2 S^2-4\right) \left(\tilde{\Phi}_e^2+\tilde{\Phi}_m^2\right)}{\left(\kappa ^2 S^2+12\right)^2}+\pi  C \left(12-5 \kappa ^2 S^2\right)-21 \kappa ^2 S^3+36 S}{48 \pi ^{3/2}  \sqrt{l^2 S C}}
\end{equation}
\begin{equation}
\small
    \mu=\frac{S \left(3 S \left(\kappa ^4 S^4+8 \kappa ^2 S^2-48\right)-\pi  C \left(144 \left(l^2 \left(\tilde{\Phi}_e^2+\tilde{\Phi}_m^2\right)-1\right)+\kappa ^4 S^4\right)\right)}{48 \pi ^{3/2}  \left(\kappa ^2 S^2+12\right) \sqrt{l^2SC^3}}
\end{equation}
Using the equation \eqref{eq:differential} on the $T$ the extremal values are given as:-
\begin{equation}
\small
    T_{ext}=\frac{-21 \kappa ^2 A_6^3+36 A_6+C \pi  \left(12-5 \kappa ^2 A_6^2\right)+\frac{432 C l^2 \pi  \left(\tilde{\Phi}_e^2+\tilde{\Phi}_m^2\right) \left(\kappa ^2 A_6^2-4\right)}{\left(\kappa ^2 A_6^2+12\right)^2}}{48  \pi ^{3/2} \sqrt{Cl^2 A_6}}
\end{equation}
\begin{equation}
    S =A_7
\end{equation}
\begin{equation}
\scriptsize
\mu_{ext}=\frac{\sqrt{A_6} \left(3 A_6 \left(\kappa ^4 A_6^4+8 \kappa ^2 A_6^2-48\right)-C \pi  \left(\kappa ^4 A_6^4+144 \left(l^2 \left(\tilde{\Phi}_e^2+\tilde{\Phi}_m^2\right)-1\right)\right)\right)}{48 C^{3/2} l \pi ^{3/2} \left(\kappa ^2 A_6^2+12\right)}
\end{equation}
Using these extremal values and using relative variables the EOS and $\tilde{m}=\mu/\mu_{ext}$ expression are given as:-
\begin{equation}
\small
   t=\frac{ \left(-21 s^3 \kappa ^2 A_6^3+36 s A_6+C \pi  \left(12-5 s^2 \kappa ^2 A_6^2\right)+\frac{432 C l^2 \pi  \left(\tilde{\Phi}_e^2+\tilde{\Phi}_m^2\right) \left(s^2 \kappa ^2 A_6^2-4\right)}{\left(s^2 \kappa ^2 A_6^2+12\right)^2}\right)}{\sqrt{s} \left(-21 \kappa ^2 A_6^3+36 A_6+C \pi  \left(12-5 \kappa ^2 A_6^2\right)+\frac{432 C l^2 \pi  \left(\tilde{\Phi}_e^2+\tilde{\Phi}_m^2\right) \left(\kappa ^2 A_6^2-4\right)}{\left(\kappa ^2 A_6^2+12\right)^2}\right)} 
\end{equation}
\begin{figure}[htp]
    \centering
    \begin{subfigure}[b]{0.2\textwidth}
        \centering
        \includegraphics[width=\textwidth]{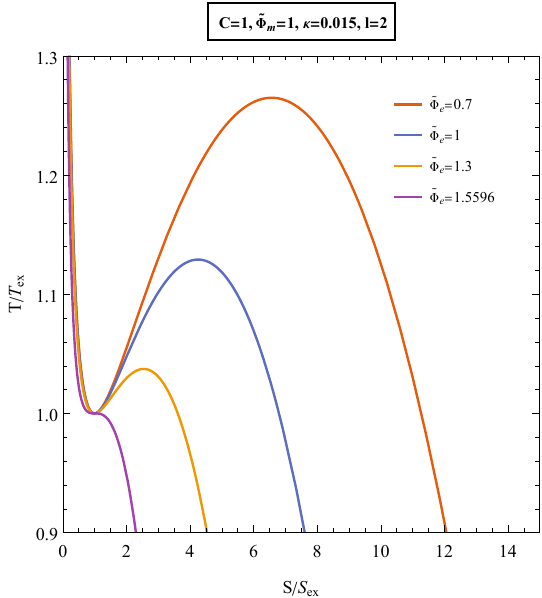} 
        \caption{}
        \label{fig:figure25}
    \end{subfigure}
    \begin{subfigure}[b]{0.2\textwidth}
        \centering
        \includegraphics[width=\textwidth]{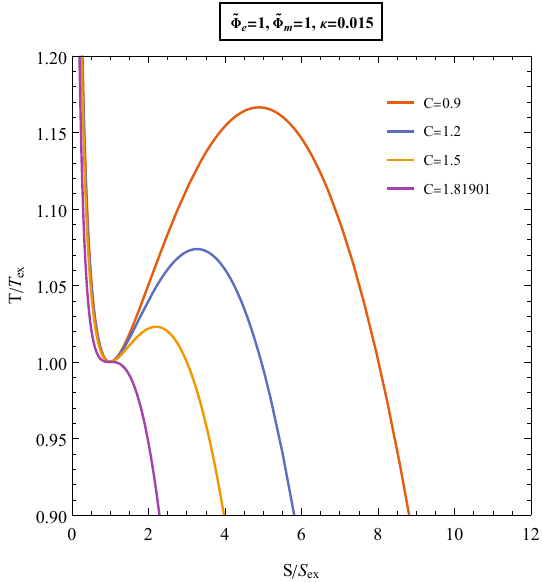} 
        \caption{}
        \label{fig:figure26}
    \end{subfigure}
    \begin{subfigure}[b]{0.2\textwidth}
        \centering
        \includegraphics[width=\textwidth]{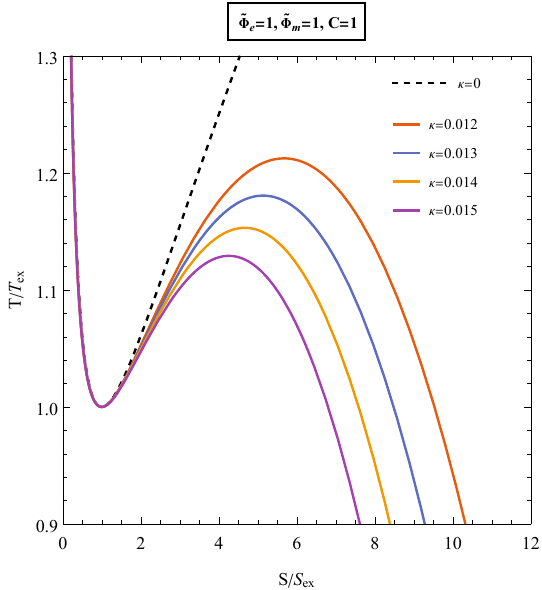} 
        \caption{}
        \label{fig:figure27}
    \end{subfigure}
    \caption{$T-S$ plots}
    \label{fig:nine}
    \end{figure}
\begin{equation}
\scriptsize
    \tilde{m}=\frac{\sqrt{s}  \left(\kappa ^2 A_6^2+12\right) \left(3 s A_6\left(s^4 \kappa ^4A_6^4+8 s^2 \kappa ^2 A_6^2-48\right)-C \pi  \left(s^4 \kappa ^4 A_6^4+144 \left(l^2 \left(\tilde{\Phi}_e^2+\tilde{\Phi}_m^2\right)-1\right)\right)\right)}{\left(s^2 \kappa ^2 A_6^2+12\right) \left(3 A_6\left(\kappa ^4 A_6^4+8 \kappa ^2 A_6^2-48\right)-C \pi  \left(\kappa ^4 A_6^4+144 \left(l^2 \left(\tilde{\Phi}_e^2+\tilde{\Phi}_m^2\right)-1\right)\right)\right)}
\end{equation}
\begin{figure}[htp]
    \centering
    \begin{subfigure}[b]{0.2\textwidth}
        \centering
        \includegraphics[width=\textwidth]{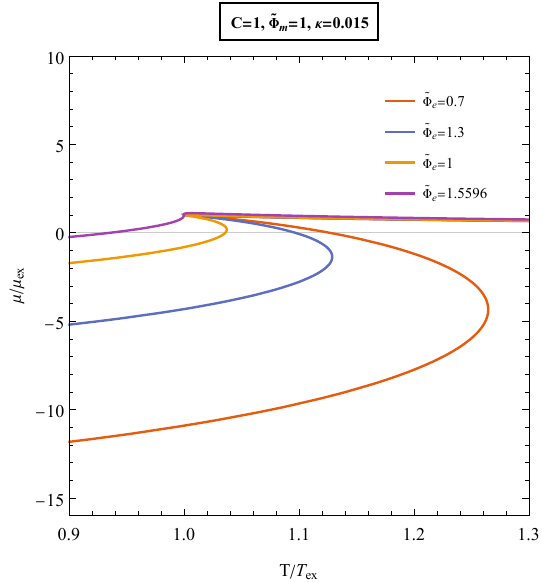} 
        \caption{}
        \label{fig:figure28}
    \end{subfigure}
    \begin{subfigure}[b]{0.2\textwidth}
        \centering
        \includegraphics[width=\textwidth]{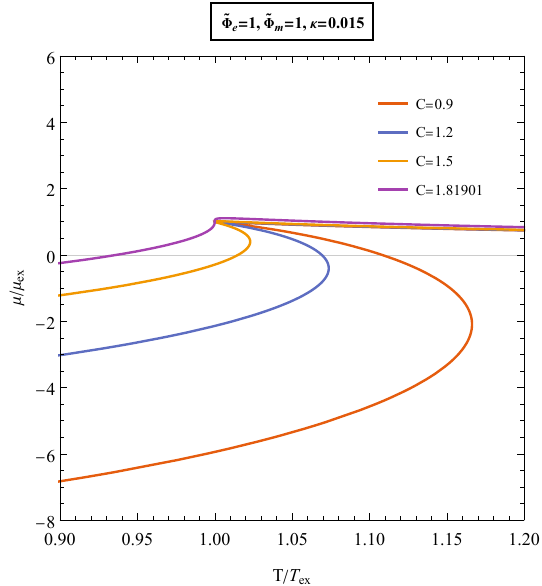} 
        \caption{}
        \label{fig:figure29}
    \end{subfigure}
    \begin{subfigure}[b]{0.2\textwidth}
        \centering
        \includegraphics[width=\textwidth]{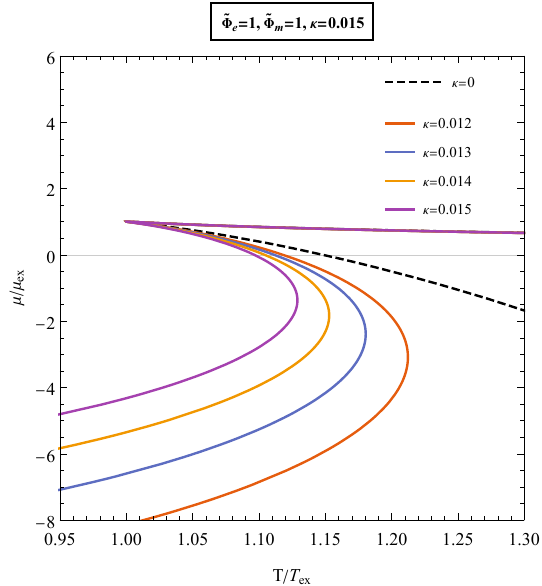} 
        \caption{}
        \label{fig:figure30}
    \end{subfigure}
    \caption{$F-T$ plots}
    \label{fig:ten}
    \end{figure}
    We plot the $T-S$ and $\mu-T$ for the various values seen in figure \ref{fig:nine}, \ref{fig:ten}. In figure \ref{fig:figure25}, \ref{fig:figure28}, we plot for the various values of $\tilde{\phi}_e$ where we see three branches namely the small (unstable), large (stable) and ultra-large (unstable) black hole branches but at a critical value of $\tilde{\phi}_e=1.5596$ the large black hole (stable) branch is missing and only two branch can be seen. We also plot for the various values of $C$ in figure \ref{fig:figure26}, \ref{fig:figure29} and we see the same kind of transitions and branches as the plots with various values of $\tilde{\phi}_e$. For the figure \ref{fig:figure27}, \ref{fig:figure28}, we plot for the various deformation parameter $\kappa$ where we see when $\kappa \rightarrow{0}$ then only two branches exist namely the small black hole and the large black hole branch but as the $\kappa$ parameter increases, we see the addition of the ultra-large black hole branch. \\
    \subsection{Other processes}
\begin{figure}[htp]
    \centering
    \begin{subfigure}[b]{0.2\textwidth}
        \centering
\includegraphics[width=\textwidth]{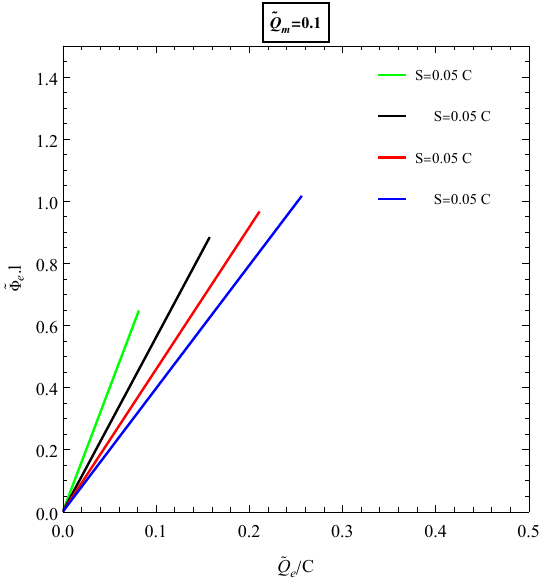} 
        \caption{}
        \label{fig:figure31}
    \end{subfigure}
     \begin{subfigure}[b]{0.2\textwidth}
        \centering
\includegraphics[width=\textwidth]{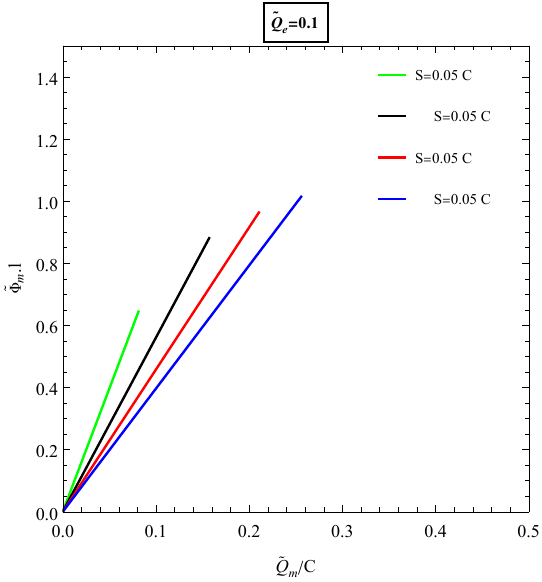} 
        \caption{}
        \label{fig:figure32}
    \end{subfigure}
    \caption{$\mu-C$ plot}
    \label{fig:eleven}
    \end{figure}

    The $\tilde{\Phi}_e-\tilde{Q}_e$ and $\tilde{\Phi}_m-\tilde{Q}_m$ processes can be studied but if we look at the equation \eqref{eq:electriccharge}, we see that the electric and the magnetic charges are directly proportional to its potentials thereby showing no forms of criticality or points or turning points. The plots are given in fig. \ref{fig:eleven}.\\
We can also study the $\mu-C$ process to see the dependency on the change of $C$. The chemical potential $\mu$ is given as:-
\begin{equation}
\small
    \mu=\frac{\pi  C S\left(12-\kappa ^2 S^2\right)-\pi ^2 \left(\tilde{Q}_e^2+\tilde{Q}_m^2\right) \left(\kappa ^2 S^2+12\right)+3 S^2 \left(\kappa ^2 S^2-4\right)}{48 \pi ^{3/2}  \sqrt{l^2 S C^2}}
\end{equation}
\begin{figure}[htp]
    \centering
    \begin{subfigure}[b]{0.2\textwidth}
        \centering
        \includegraphics[width=\textwidth]{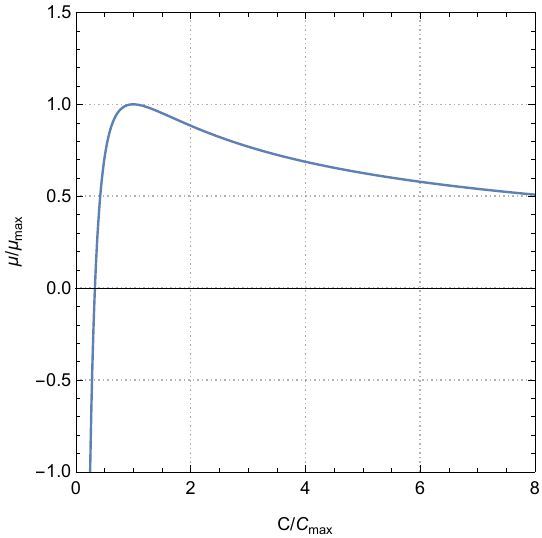} 
        \caption{}
        \label{fig:figure33}
    \end{subfigure}
    \caption{$\mu-C$ plot}
    \label{fig:twelve}
    \end{figure}
We calculate the point of extremity and write it as:-
\begin{equation}
   C_{max}=\frac{9 S^2 \left(\kappa ^2 S^2-4\right)-3 \pi ^2 \left(\tilde{Q}_e^2+\tilde{Q}_m^2\right) \left(\kappa ^2 S^2+12\right)}{\pi  S \left(\kappa ^2 S^2-12\right)} 
\end{equation}
\begin{equation}
\small
    \mu_{max}=\frac{\left(-\kappa ^2 S^2+12\right) }{72 l^2}\sqrt{\frac{l^2 S^2 \left(12-\kappa ^2 S^2\right)}{3 \pi ^2 \left(\tilde{Q}_e^2+\tilde{Q}_m^2\right) \left(\kappa ^2 S^2+12\right)-9 \kappa ^2 S^4+36 S^2}}
\end{equation}
Using the dimensional parameters $c=C/C_{max}$, $m=\mu/\mu_{max}$ we get:-
\begin{equation}
\label{eq:muc}
   m= \frac{(3 c-1)\sqrt{c}}{2 c^2 }
\end{equation}
From fig \ref{fig:twelve}, we see the $\mu-C$ plot which is plotted using the eq. \eqref{eq:muc} which is only dependendant on the $C$. The plot is the same as studied for the different black hole systems studied and also for different entropy models studied using the RPST formalism.

\section{Discussion}
\label{sec:Discussions}
We study the Restricted Phase Space (RPS) Thermodynamics of charged AdS black holes in the context of Kaniadakis statistics rather than the traditional Gibbs-Boltzmann statistics. in Section \ref{sec:introduction} we give a brief introduction about the emergence of black hole thermodynamics study with the introduction of Hawking's temperature $T$ and the Bekenstein-Hawking entropy $S$ and how it was enhanced by the introduction of EPST and Visser's formalism. Still, due to certain issues like the ensemble ambiguity due to cosmological constant variation, we use a stricter version of Visser's formalism named RPST. We also give a brief idea of the richness of the magnetic monopole introduction known as a dyonic black hole and how it gave way to various studies. In Section \ref{sec:RPST} We start the RPS study by calculating the mass from the line element and then adding rescaled charges and also the central charge $C$. In Section \ref{sec:EOS} we incorporate the Kaniadakis entropy using the usual method and calculated various parameters. By studying various $T-S$ and $F-T$ processes we can state many things. In Figure \ref{fig:figure1} and \ref{fig:figure5} we see at least 3 types of phase transition. First is the Van der Waals phase transition of the first order at supercritical values of $q_e=\tilde{Q}_e/\tilde{Q}_e^c$ then at the critical value, we see a second order phase transition at the critical value and at a certain supercritical value we also see non-equilibrium phase transition which is also known as the Hawking-Page phase transition. In all the plots we see an extra branch at higher entropy due to the addition of the Kaniadakis deformation parameter $\kappa$. Rest plots \ref{fig:figure2}, \ref{fig:figure3}, \ref{fig:figure6}, \ref{fig:figure7} show the variation of the $C$ and $\kappa$ where we see a similarity in its phase plots hinting a connection between the deformation parameter and the $C$ which gives the total number of degrees of freedom or the information at the boundary. When the magnetic charge is turned off seen in figure \ref{fig:figure9}, \ref{fig:figure12} we see the Hawking-page phase transition vanish but now we see the Van der Waals phase transition for the subcritical values of the electric charge $\tilde{Q}_e$. We also see here the extra brach due to the deformation parameter. When both the charges are turned off it forms towards the Schwarzschild AdS black hole with the extra branch of unstable ultra-large black hole seen in figure \ref{fig:figure15} and \ref{fig:figure16}. When we study for the electric potential and magnetic charge we see a novel behavior that was not seen for dyonic black holes. We see a superfluid $\lambda$ phase transition which matches with the condensed matter systems as the fluid/superfluid transitions. We think that this was not seen before as this was possible due to the different ensemble studies taken into consideration thereby one can say that phase transition is highly dependent on the choice of ensemble. We see in  We see in \ref{fig:figure231}, \ref{fig:figure251} we do see a $\lambda$ phase transition but due to the deformation parameter, it does come out properly. Only at a particular $\tilde{Q}_m$ do we see the superfluid $\lambda $ phase transition. Setting both potentials we see that plots are similar to the plots when both the charges are turned off. Since the charges are proportional to the potentials so we do not see any kind of criticality or point of inflection when checking $\tilde{\Phi}_e-\tilde{Q}_e$ and $\tilde{\Phi}_m-\tilde{Q}_m$ processes. Lastly, for the $\mu-C $ process we see a similar kind of branch seen in various types of black hole studied under RPST and various entropy models.
\newpage
\newpage
\appendix

\section{}
\begin{equation}
\scriptsize
    A_1=\text{Root}\left[5 \pi  \text{$\#$1}^4 C \kappa ^4+176 \pi  \text{$\#$1}^2 C \kappa ^2+70 \text{$\#$1}^5 \kappa ^4+1680 \text{$\#$1}^3 \kappa ^2-288 \text{$\#$1}+48 \pi  C\&,2\right]
\end{equation}
\begin{equation}
\scriptsize
    A_2=\sqrt[3]{-125 \pi ^3 C^3 \kappa ^6+210 \sqrt{15} \sqrt{\kappa ^6 \left(25 \pi ^4 C^4 \kappa ^4+8460 \pi ^2 C^2 \kappa ^2-3024\right)}-75600 \pi  C \kappa ^4}
\end{equation}
\begin{equation}
\scriptsize
    A_3=\sqrt[3]{-25 \pi ^3 C^3 \kappa ^6+42 \sqrt{15} \sqrt{\kappa ^6 \left(25 \pi ^4 C^4 \kappa ^4+8460 \pi ^2 C^2 \kappa ^2-3024\right)}-15120 \pi  C \kappa ^4}
\end{equation}
\begin{equation}
\scriptsize
\begin{split}
&    A_4=\text{Root}\left[875 \kappa ^8 \text{$\#$1}^{10}+100 C \pi  \kappa ^8 \text{$\#$1}^9+\left(-15 \pi ^2 \tilde{Q}_m^2 \kappa ^8-19920 \kappa ^6\right) \text{$\#$1}^8-2720 c \pi  \kappa ^6 \text{$\#$1}^7\right.\\
&    \left.+\left(480 \pi ^2 \tilde{Q}_m^2 \kappa ^6-38880 \kappa ^4\right) \text{$\#$1}^6-4032 C \pi  \kappa ^4 \text{$\#$1}^5+\left(5664 \pi ^2 \tilde{Q}_m^2 \kappa ^4-66816 \kappa ^2\right) \text{$\#$1}^4\right.\\
&\left.+\left(6912-27648 \pi ^2 \tilde{Q}_m^2 \kappa ^2\right) \text{$\#$1}^2-6912 \pi ^2 \tilde{Q}_m^2\&,4\right]
    \end{split}
\end{equation}
\begin{equation}
\scriptsize
\begin{split}
  &  A_5=\text{Root}\left[100 \pi  \text{$\#$1}^9 C \kappa ^8-2720 \pi  \text{$\#$1}^7 C \kappa ^6-4032 \pi  \text{$\#$1}^5 C \kappa ^4+875 \text{$\#$1}^{10} \kappa ^8\right.\\
&   \left. +\text{$\#$1}^8 \left(-19920 \kappa ^6-15 \pi ^2 \kappa ^8 \tilde{Q}_m^2\right)+\text{$\#$1}^6 \left(480 \pi ^2 \kappa ^6 \tilde{Q}_m^2-38880 \kappa ^4\right)+\right.\\
&\left.\text{$\#$1}^4 \left(5664 \pi ^2 \kappa ^4 \tilde{Q}_m^2-66816 \kappa ^2\right)+\text{$\#$1}^2 \left(6912-27648 \pi ^2 \kappa ^2 \tilde{Q}_m^2\right)-6912 \pi ^2 \tilde{Q}_m^2\&,4\right]
    \end{split}
\end{equation}
\begin{equation}
\scriptsize
    B_1=\left(\frac{5 A_2}{\kappa ^2}+\frac{5\ 5^{2/3} \left(5 \pi ^2 C^2 \kappa ^2+252\right)}{A_3}-25 \pi  C\right)
\end{equation}
\begin{equation}
\scriptsize
\begin{split}
 &   A_6=\text{Root}\left[35 \kappa ^8 \text{$\#$1}^9+5 C \pi  \kappa ^8 \text{$\#$1}^8+1248 \kappa ^6 \text{$\#$1}^7+184 C \pi  \kappa ^6 \text{$\#$1}^6+14688 \kappa ^4 \text{$\#$1}^5-6912 C l^2 \pi  \tilde{\Phi}_m^2\right.\\
 &\left.+\left(720 C l^2 \pi  \tilde{\Phi}_e^2 \kappa ^4+720 C l^2 \pi  \tilde{\Phi}_m^2 \kappa ^4+2304 C \pi  \kappa ^4\right) \text{$\#$1}^4+55296 \kappa ^2 \text{$\#$1}^3-20736 \text{$\#$1}\right.\\
 &\left.+\left(-10368 C l^2 \pi  \tilde{\Phi}_e^2 \kappa ^2-10368 C l^2 \pi  \tilde{\Phi}_m^2 \kappa ^2+10368 C \pi  \kappa ^2\right) \text{$\#$1}^2-6912 C l^2 \pi  \tilde{\Phi}_e^2+6912 C\pi \&,2\right]
    \end{split}
\end{equation}
\begin{equation}
\scriptsize
\begin{split}
 &   A_7= \text{Root}\left[5 \pi  \text{$\#$1}^8 C \kappa ^8+184 \pi  \text{$\#$1}^6 C \kappa ^6+\text{$\#$1}^4 \left(2304 \pi  C \kappa ^4+720 \pi  C \kappa ^4 l^2\tilde{\Phi}_e^2+720 \pi  C \kappa ^4 l^2 \tilde{\Phi}_m^2\right)\right.\\
 &\left.+\text{$\#$1}^2 \left(10368 \pi  C \kappa ^2-10368 \pi  C \kappa ^2 l^2 \tilde{\Phi}_e^2-10368 \pi  C \kappa ^2 l^2 \tilde{\Phi}_m^2\right)+35 \text{$\#$1}^9 \kappa ^8+1248 \text{$\#$1}^7 \kappa ^6\right.\\
 &\left.+14688 \text{$\#$1}^5 \kappa ^4+55296 \text{$\#$1}^3 \kappa ^2-20736 \text{$\#$1}-6912 \pi  C l^2 \tilde{\Phi}_e^2-6912 \pi  C l^2 \tilde{\Phi}_m^2+6912 \pi  C\&,2\right]
    \end{split}
\end{equation}
\begin{equation}
\scriptsize
D_1=\left(\tilde{Q}_m^2+\frac{q_e^2 A_1^3 \left(-\frac{12 \pi ^2 \tilde{Q}_m^2}{A_1^3}+5 C \pi  \kappa ^2+35 \kappa ^2 A-1-\frac{\pi ^2 \tilde{Q}_m^2 \kappa ^2+12}{A_1}+\frac{4 C \pi }{A_1^2}\right)}{\pi ^2 \left(\kappa ^2 A_1^2+12\right)}\right)
\end{equation}
\begin{equation}
\scriptsize
\begin{split}
  &  D_2=4 C \pi  \left(9 s^3-q_e^2 s^2+3 s+45 q_e^2\right) \kappa ^2 A-1^3+12 \left(\left(\pi ^2 \tilde{Q}_m^2 \left(q_e^2-1\right) \kappa ^2-12\right) s^2\right.\\
  & \left. +3 \pi ^2 \tilde{Q}_m^2 \kappa ^2-3 \text{$\mathit{q}$e}^2 \left(\pi ^2 \tilde{Q}_m^2 \kappa ^2+12\right)\right) A_1^2+144 C \pi  \left(q_e^2+s\right) A-1-432 \pi ^2 \tilde{Q}_m^2 \left(q_e^2-1\right)
    \end{split}
\end{equation}
\begin{equation}
\scriptsize
    D_3=\frac{3 A_1^3 \left(35 \kappa ^2 A_1-\frac{12}{A_1}+\frac{4 \pi  C}{A_1^2}+5 \pi  C \kappa ^2\right) \left(\kappa ^2 A_1^2-4\right)}{\kappa ^2 A_1^2+12}
\end{equation}
\begin{equation}
\scriptsize
\begin{split}
&    D_4=5 \sqrt{\frac{l^2 A_1}{C}} \left(21 \kappa ^4 s^2 \left(s^2-5 q_e^2\right) A_1^5+5 \pi  C \kappa ^4 s^2 \left(\text{sk}-3 q_e^2\right) A_1^4+12 \kappa ^2 \left(35 q_e^2\right.\right.\\
&\left.\left.+3 \left(q_e^2-1\right) s^2+21 s^4\right) A_1^3+12 \pi  C \kappa ^2 \left(5 q_e^2-q_e^2 s^2+5 s^3-s\right) A_1^2\right.\\
&\left.-144 \left(q_e^2+3 s^2\right)A_1+48 \pi  C \left(q_e^2-3 s\right)\right)
    \end{split}
\end{equation}
\begin{equation}
\scriptsize
    D_5=4 s \left(5 \pi  C \kappa ^4 A_1^4-3360 \kappa ^2 A_1^3-384 \pi  C \kappa ^2 A_1^2+1152 A_1+48 \pi  C\right) \sqrt{\frac{l^2 s A_1}{C}}
\end{equation}
\begin{equation}
\scriptsize
\begin{split}
   & D_6=7  \left(5 \kappa ^4 s^2 \left(3 s^2-7 q_e^2\right) A_1^5+\pi  C \kappa ^4 s^2 \left(3 s-5 q_e^2\right) A_1^4+12 \kappa ^2 \left(105 q_e^2\right.\right.\\
   & \left.\left.+\left(q_e^2-1\right) s^2+15 s^4\right) A_1^3+4 \pi  C \kappa ^2 \left(45 q_e^2-q_e^2 \text{sk}^2+9 s^3+3 s\right) A_1^2\right.\\
   &\left.-144 \left(3 q_e^2+s^2\right) A_1+144 \pi  C \left(q_e^2+s\right)\right)
    \end{split}
\end{equation}
\begin{equation}
\scriptsize
\begin{split}
  &  D_7=l^2 \left(-105 \kappa ^6 A_4^8-20 C \pi  \kappa ^6 A_4^7+\left(9 \pi ^2 \tilde{Q}_m^2 \kappa ^6+444 \kappa ^4\right) A_4^6+144 C \pi  \kappa ^4 A_4^5\right.\\
  &\left.-36 \kappa ^2 \left(3 \pi ^2 \tilde{Q}_m^2 \kappa ^2-52\right) A_4^4+432 \left(\pi ^2 \tilde{Q}_m^2 \kappa ^2-4\right) A_4^2-576 \pi ^2 \tilde{Q}_m^2\right)
    \end{split}
\end{equation}
\begin{equation}
\scriptsize
\begin{split}
   & D_8=\left(5 \kappa ^4 A_4^4-72 \kappa ^2 A_4^2-48\right) \left(-21 s^4 \kappa ^2 A_4^4+36 s^2 A_4^2+C \pi  s \left(12-5 s^2 \kappa ^2 A_4^2\right) A_4\right.\\
   &\left.+3 \pi ^2 \left(s^2 \kappa ^2 A_4^2-4\right) \left(\tilde{Q}_m^2\right.\right.\\
   &\left.\left.+\frac{s \tilde{\Phi}_e^2 \left(\kappa ^2 A_4^2+12\right)^3 \left(-35 \kappa ^2 A_4^4-5 C \pi  \kappa ^2 A_4^3+\left(\pi ^2 \tilde{Q}_m^2 \kappa ^2+12\right) A_4^2-4 C \pi  A_4+12 \pi ^2 \tilde{Q}_m^2\right)}{\pi ^2 \left(s^2 \kappa ^2 A_4^2+12\right)^2 \left(5 \kappa ^4 A_4^4-72 \kappa ^2 A_4^2-48\right)}\right)\right)
    \end{split}
\end{equation}
\begin{equation}
\scriptsize
\begin{split}
&    D_9=2 s^{3/2} \left(-105 \kappa ^6 A_4^8-20 C \pi  \kappa ^6 A_4^7+\left(9 \pi ^2 \tilde{Q}_m^2 \kappa ^6+444 \kappa ^4\right) A_4^6+144 C \pi  \kappa ^4 A_4^5\right.\\
&\left.-36 \kappa ^2 \left(3 \pi ^2 \tilde{Q}_m^2 \kappa ^2-52\right)A_4^4+432 \left(\pi ^2 \tilde{Q}_m^2 \kappa ^2-4\right) A_4^2-576 \pi ^2 \tilde{Q}_m^2\right)
\end{split}
\end{equation}
\begin{equation}
\scriptsize
\begin{split}
 &   D_{10}=25 \kappa ^6 A_4^8-3 \kappa ^4 \left(\pi ^2 \tilde{Q}_m^2 \kappa ^2-92\right) A-4^6+128 C \pi  \kappa ^4 A_4^5-12 \kappa ^2 \left(\pi ^2 \tilde{Q}_m^2 \kappa ^2-228\right) A_4^4\\
 &+48 \left(5 \pi ^2 \tilde{Q}_m^2 \kappa ^2-12\right) A_4^2-576 \pi ^2 \tilde{Q}_m^2
 \end{split}
\end{equation}
\begin{equation}
\scriptsize
\begin{split}
  &  D_{11}= \left(5 \kappa ^4 A_4^4-72 \kappa ^2 A_4^2-48\right) \left(-C \pi  s^5 \kappa ^4 A_4^5+3 s^2 \left(s^4 \kappa ^4 A_4^4+8 s^2 \kappa ^2 A_4^2-48\right) A_4^2\right.\\
  &\left.+144 C \pi  s A_4-\pi ^2\tilde{Q}_m^2 \left(s^2 \kappa ^2 A_4^2+12\right)^2\right.\\
  &\left.+\frac{s \tilde{\Phi}_e^2 \left(\kappa ^2 A_4^2+12\right)^3 \left(35 \kappa ^2 A_4^4+5 C \pi  \kappa ^2 A_4^3-\left(\pi ^2 \tilde{Q}_m^2 \kappa ^2+12\right) A_4^2+4 C \pi  A_4-12 \pi ^2 \tilde{Q}_m^2\right)}{5 \kappa ^4A_4^4-72 \kappa ^2 A_4^2-48}\right)
    \end{split}
\end{equation}
\begin{equation}
\scriptsize
\begin{split}
&    D_{12}=2 \sqrt{s} \left(s^2 \kappa ^2 A_4^2+12\right) \left(25 \kappa ^6 A_4^8-3 \kappa ^4 \left(\pi ^2 \tilde{Q}_m^2 \kappa ^2-92\right) A_4^6+128 C \pi  \kappa ^4 A_4^5\right.\\
&\left.-12 \kappa ^2 \left(\pi ^2 \tilde{Q}_m^2 \kappa ^2-228\right) A_4^4+48 \left(5 \pi ^2 \tilde{Q}_m^2 \kappa ^2-12\right) A_4^2-576 \pi ^2 \tilde{Q}_m^2\right)
    \end{split}
\end{equation}

\begin{equation}
\begin{split}
&t_1=5 \left(\kappa ^2 A_1^2+12\right) \left(-21 s^4 \kappa ^2 A_1^4+36 s^2 A_1^2+C \pi  s \left(12-5 s^2 \kappa ^2 A\right.\right.\\
&\left.\left.-1^2\right) A-1+3 \pi ^2 \left(s^2 \kappa ^2 A_1^2-4\right) D_1\right)\\
&t_2=4 C \left(\frac{  A_1}{C}\right)s^{3/2} \left(5 C \pi  \kappa ^4 A_1^4-3360 \kappa ^2 A_1^3-384 C \pi  \kappa ^2 A_1^2\right.\\&\left.+1152 A_1+48 C \pi \right)
\end{split}
\end{equation}

\begin{equation}
\begin{split}
& f_1=7 l^2 \left(5 s^2 \left(3 s^2-7 \text{$\mathit{q}$e}^2\right) \kappa ^4 A_1^6+C \pi  s^2 \left(3 s-5 \text{$\mathit{q}$e}^2\right) \kappa ^4 A_1^5\right.\\
&\left.+\kappa ^2 \left(180 s^4+\left(\text{$\mathit{q}$e}^2-1\right) \left(\pi ^2 \text{qm}^2 \kappa ^2+12\right) s^2+1260 \text{$\mathit{q}$e}^2\right) A_1^4+D_2\right)\\
& f_2=4 C \sqrt{\frac{l^2 A_1}{C}} \sqrt{\frac{l^2 s A_1}{C}} \left(C \pi  \kappa ^4 A_1^4-3360 \kappa ^2 A-1^3\right.\\
& \left.-480 C \pi  \kappa ^2 A_1^2+1152 A_1-528 C \pi \right)
\end{split}
\end{equation}

\end{document}